\documentclass[superscriptaddress,floatfix,nofootinbib,12pt,prx]{revtex4}
\usepackage{graphicx,amsfonts,amssymb,amsmath,hyperref,hypcap,enumerate}
\usepackage{slashed}
\usepackage{amsmath,amssymb,bm,graphicx,dsfont}
\usepackage[latin9]{inputenc}
\usepackage{amsmath}
\usepackage{slashed}
\usepackage{dsfont}
\usepackage{amstext}
\usepackage{amssymb}
\usepackage{amsbsy}
\usepackage{amsthm}
\usepackage{epsfig}
\usepackage{graphicx}
\usepackage{bbm}
\usepackage{hyperref}
\usepackage{color}
\usepackage{slashed}
\renewcommand{\d}{\partial}

\newcommand{\be}{\begin{equation}}
\newcommand{\ee}{\end{equation}}
\newcommand{\bea}{\begin{eqnarray}}
\newcommand{\eea}{\end{eqnarray}}
\newcommand{\beq}{\begin{equation}}
\newcommand{\eeq}{\end{equation}}
\newcommand{\beqn}{\begin{eqnarray}}
\newcommand{\eeqn}{\end{eqnarray}}

\renewcommand{\hat}[1]{{\widehat #1}}

\def\nn{\nonumber}

\begin{document}
\title{Intrinsic and emergent anomalies at deconfined critical points}
\author{Max A. Metlitski}
\affiliation{Department of Physics, Massachusetts Institute of
Technology, Cambridge, MA 02139, USA}

\author{Ryan Thorngren}
\affiliation{Department of Mathematics, University of California,
Berkeley, California 94720, USA}

\date{\today}

\begin{abstract}
It is well known that theorems of Lieb-Schultz-Mattis type prohibit the existence of a trivial symmetric gapped ground state in certain systems possessing a combination of internal and lattice symmetries. In the continuum description of such systems the Lieb-Schultz-Mattis theorem is manifested in the form of a quantum anomaly afflicting the symmetry.  We demonstrate this phenomenon in the context of the deconfined critical point between a Neel state and a valence bond solid in an $S  =1/2$  square lattice antiferromagnet, and compare it to the case of $S=1/2$ honeycomb lattice where no anomaly is present. We also point out that new anomalies, unrelated to the microscopic Lieb-Schultz-Mattis theorem, can emerge prohibiting the existence of a trivial gapped state in the immediate vicinity of  critical points or phases. For instance, no translationally invariant weak perturbation of the $S = 1/2$ gapless spin chain can open up a trivial gap even if the spin-rotation symmetry is explicitly broken. The same result holds for the $S  =1/2$ deconfined critical point on a square lattice.


\noindent

\end{abstract}

\maketitle

\section{Introduction}

The Lieb-Schultz-Mattis (LSM) theorem\cite{LSM} and its generalization to higher dimensions\cite{Hastings, Oshikawa} states that an insulator with half-odd-integer spin  per unit cell cannot 
have a trivial gapped ground state: in 1+1D the  ground state must either break the translational symmetry or be gapless, while in higher dimensions the system may also spontaneously break the $SO(3)_s$ spin rotation symmetry or support topological order. In  recent  years, this result has been generalized to a variety of cases where one relies on lattice symmetries other than translation - e.g. rotation, reflection or glide - in combination with $SO(3)_s$, or replaces $SO(3)_s$ by time-reversal symmetry, to rule out a trivial gap.\cite{MikeHaruki2015,MikeHaruki, FuLMS, LuConstr, LuRanOshikawa, VishwanathRan} Furthermore, it was noted that 
the impossibility of a trivial gap  is very reminiscent of the situation occurring on the boundary of a topological insulator, or a more general symmetry protected topological (SPT) phase. In fact, one may view a system with $S = 1/2$ per unit cell as a boundary of a crystalline SPT phase protected by a combination of translational symmetry and $SO(3)_s$.\cite{BarkeshliCryst} Such a crystalline SPT can be constructed as an array of 1+1D Haldane chains - then the boundary is an array of ``dangling" spin-$1/2$'s.  As we will see, the higher-dimensional bulk is a useful ``conceptual" tool, even in cases when it is physically absent. 

For SPT phases protected just by internal symmetry the relationship between the bulk topological invariant and the non-triviality of the surface is very-well understood - the boundary realizes the symmetry in a non-onsite manner. If one attempts to gauge the symmetry in the boundary theory, one runs into an inconsistency - an anomaly. This anomaly is, however, cured by the bulk of the system. This means that every surface phase, no matter whether it is symmetry broken, gapless or topologically ordered, must realize the same anomaly which matches the bulk - a property that must be implemented by the low-energy continuum theory describing each surface phase. What about the  the bulk/boundary relationship for a crystalline SPT protected by a combination of lattice and internal symmetries or equivalently, how do LSM constraints enter in the low-energy continuum theory? Here, we  discuss two examples: i) the gapless $S  =1/2$ spin chain in 1+1D; ii) the deconfined quantum critical point (QCP) in 2+1D between an $S=1/2$ Neel state and a valence bond solid (VBS) on square and honeycomb lattices.\cite{deccp, deccplong} In these examples, we focus on the following symmetries:  $SO(3)_s$, translations and (in 2+1D) lattice rotations. We find that the LSM-like anomaly may be determined by treating the lattice symmetries in the low-energy theory as {\it internal} symmetries. In the case of rotations, this is done by combining the microscopic rotation symmetry with the emergent Lorentz symmetry of the continuum field theory. In particular, we find that for the $S = 1/2$ square lattice the combination of $SO(3)_s$ and translations is anomalous, and also the combination of $SO(3)_s$ and 180 degree rotations is anomalous. This is in complete agreement with LSM-like theorems.\cite{MikeHaruki}  On the other hand, on the honeycomb lattice, we find no anomalies for the symmetries listed above. Again, this is consistent since a trivial symmetric gapped state on the honeycomb lattice has been recently constructed.\cite{YRanHoney,ZaletelHoney} The treatment of lattice symmetries as internal symmetries for the purpose of anomaly computation is consistent with Ref.~\onlinecite{RyanDom}, which argues that the classification of crystalline SPTs with a symmetry group $G$ comprising both lattice and internal symmetries  is identical to the classification of SPTs with a purely internal symmetry group $G$ (see also Ref.~\onlinecite{HermeleCryst}). It is also consistent with the results of Ref.~\onlinecite{BarkeshliCryst} obtained in the context of topologically ordered 2+1D phases with crystalline symmetries.  

In addition to the anomalies mandated by LSM-like theorems, we find that new anomalies can emerge in the neighbourhood of critical points/phases. This occurs when the microscopic symmetry group $G$ does not act on the gapless degrees of freedom in the critical theory in a faithful manner: $G$ may act as $G/H$, where $H$ is a normal subgroup. There are cases when $G/H$ has an anomaly even though $G$ itself does not.\footnote{This situation was recently discussed in Ref.~\cite{JuvenGH} where it was used to construct symmetric gapped surface states of SPT phases.} Then no $G$-symmetric infinitesimal  perturbation of the critical theory can open up a trivial gap.\footnote{We assume here that no ``accidental" strongly first order transition to the regime outside of field theory validity occurs upon adding the infinitesimal perturbation.} Physically, there are not enough degrees of freedom in the critical theory in order to drive the system into a trivial phase. However, if we perturb  the system strongly, states transforming non-trivially under $H$ may eventually come down in energy and a trivial gapped ground state may be achieved. An example of this is provided by the 1+1D $S  =1/2$ chain. Here the gapless excitations sit at points $k  =0$ and $k = \pi$ in the Brillouin zone. Therefore, the translational symmetry $Z$ acts as $Z_2$ in the continuum theory. It has long been known that this $Z_2$ symmetry is anomalous.\cite{WittenG, Oshikawa2017} What this, however, means is that no weak perturbation can gap out the $S  =1/2$ chain without breaking the translational symmetry, even if the perturbation completely breaks spin-rotations (and time-reversal). This is consistent with what we know: for instance, if we start with the isotropic antiferromagnetic Heisenberg chain and introduce a weak Ising asymmetry $\Delta H = \delta \sum_i S^z_i S^z_{i+1}$, $\delta > 0$, this drives the system into an Ising antiferromagnet, $\langle S^z_i \rangle \sim (-1)^{i}$, which spontaneously breaks the translation symmetry (the $S^z$ spin-rotation symmetry and time-reversal can be further broken with a small uniform Zeeman field). 
Other nearby gapped states, such as the VBS also break translations. Of course, if one applies a strong enough Zeeman field, one completely polarizes the chain consistent with the fact that there is no intrinsic LSM-like anomaly for translational symmetry alone. This, however, requires a critical strength of the Zeeman field and does not occur in the immediate vicinity of the gapless state.

\begin{figure}[t]
\includegraphics[width=0.5\columnwidth]{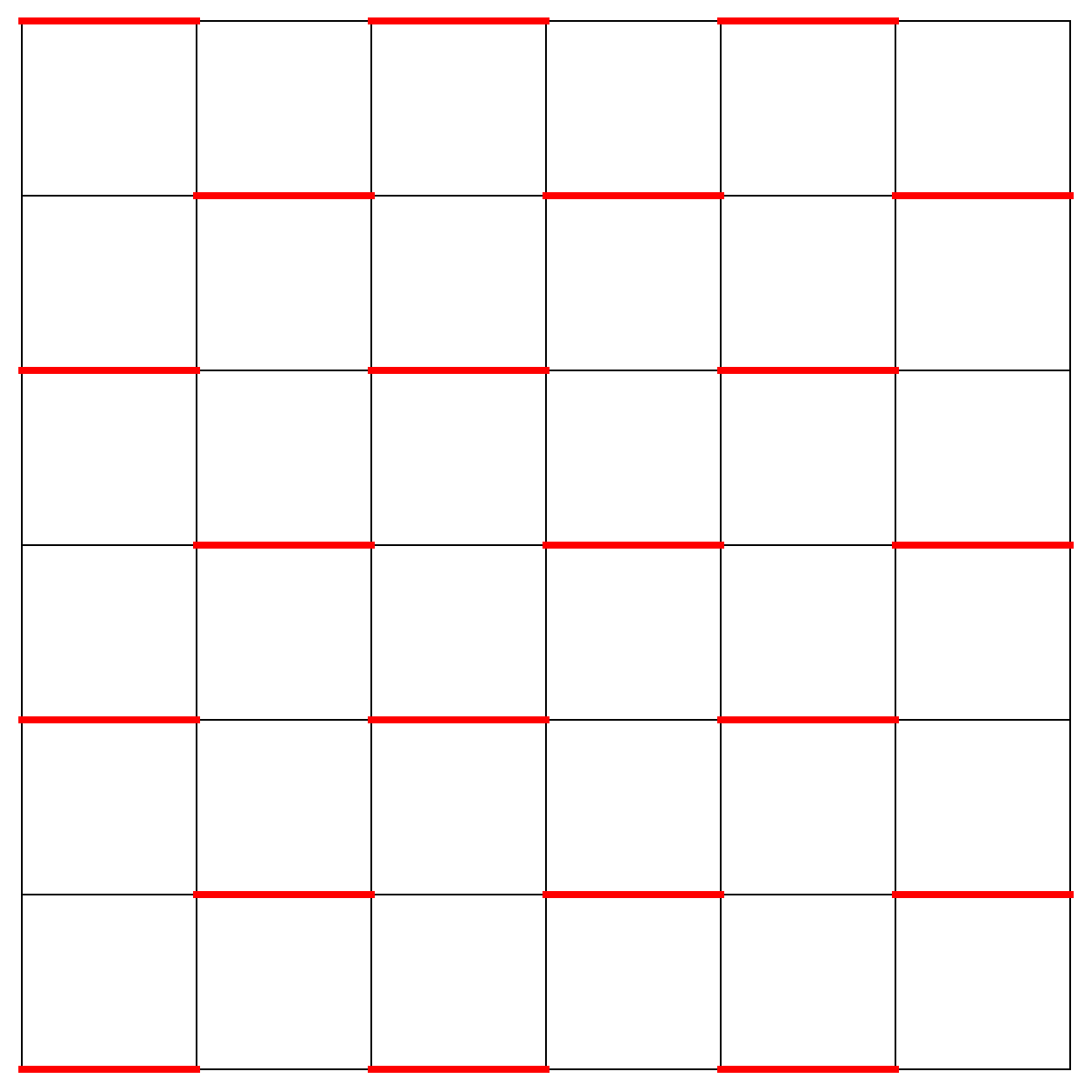}
\caption{A staggering of bond strengths for an $S  =1/2$ square lattice. Weakly perturbing the deconfined critical point with such a staggering cannot open a trivial gap, while preserving $T_xT_y$ and $SO(3)_s$ symmetry.}
\label{fig:stag}
\end{figure}

We find that similar new anomalies emerge at the deconfined critical point in an $S =1/2$ square lattice magnet. Here, the translational symmetry $Z^x \times Z^y$ acts in a $Z^x_2 \times Z^y_2$ manner on the gapless decrees of freedom at the QCP. Furthermore, we find that this $Z^x_2 \times Z^y_2$ symmetry is anomalous. Thus, again, no weak perturbation can drive the system into a translationally invariant gapped phase (even if it breaks the $SO(3)_s$ symmetry). Another emergent anomaly is present for the combination of diagonal translations $T_x T_y$ and $SO(3)_s$. While one might have naively thought that by staggering the bond-strengths as in Fig.~\ref{fig:stag} one can immediately trivially gap out the deconfined critical point, this is not the case - a finite strength of such staggering is needed for a trivial gap to open. In contrast, we find no emergent anomalies for the combination of translations, rotations and $SO(3)_s$ for an $S =1/2$ honeycomb lattice (and as already mentioned, no intrinsic LSM anomalies). This, in principle, opens the possibility that in the CP$^1$ field theory perturbed by triple monopoles governing the deconfined QCP on the honeycomb lattice an intermediate trivial symmetric phase may exist between the Neel state and the VBS state. However, current studies of lattice models on the honeycomb lattice suggest either a continuous direct transition or a weakly first-order transition.\cite{DamleH1, KawashimaH, BlockMelko, DamleH2} 
 Refs.~\onlinecite{DamleH1, DamleH2} also argue based on the anisotropy of VBS histograms that the triple monopole operator is nearly marginal at the transition - it may be that the system sizes probed in Refs.~\cite{DamleH1, KawashimaH, BlockMelko, DamleH2} were not large enough to study the true IR effects of this operator. If this operator is slightly relevant it is possible that it eventually drives the system to a trivial gapped state, opening up a narrow region of intermediate gapped phase near the putative QCP. Of course, a less exciting scenario where this operator drives a first order transition or leads to coexistence of the Neel and VBS phases is also possible. In any case, these findings motivate further numerical study of the Neel-VBS transition on the honeycomb lattice.  

We would like to point out that the situation of emergent anomalies described above should not be confused with the case when the microscopic symmetry $G$ is dynamically enlarged in the critical state to a larger group $G'$, i.e. when perturbations breaking $G'$ to $G$ are irrelevant in the RG sense. In such cases, the enlarged symmetry may also be anomalous. An example is provided by the 1+1D $S =1/2$ chain where the microscopic $SO(3)_s \times Z^x_2$ symmetry is dynamically enlarged to $SO(4)$. Similarly there is evidence that the $SO(3)_s \times \left[Z^{rot}_4 \rtimes Z^x_2\right]$ symmetry of the $S=1/2$ square lattice deconfined QCP is dynamically enlarged to an $SO(5)$ symmetry (here $Z^{rot}_4$ stands for $90$ degree rotations). The anomaly associated with this $SO(5)$ symmetry has been determined in Ref.~\onlinecite{SO5gang}, and may be used as a starting point to derive the intrinsic/emergent anomalies associated with the physical symmetries studied here.\cite{ZoharRyan} However, it is not necessary to assume this emergent $SO(5)$ either to compute the anomaly associated with the physical symmetry or to study its consequences.

In addition to the above anomaly analysis, we discuss the dynamics of the Neel-VBS transition of an $S = 1/2$ rectangular lattice and $S  =1$ square lattice. Some time ago, it was suggested  that the Neel-VBS transition of an $S  = 1/2$ rectangular lattice may be  continuous and may possess an emergent $O(4)$ symmetry.\cite{tsmpaf06} However,  numerical simulations of Ref.~\onlinecite{BlockMelko} have found a first order transition on a rectangular lattice, so this proposal was abandoned. Here, we would like to revisit this proposal in view of recent theoretical\cite{qeddual,karchtong,SO5gang} and numerical progress.\cite{emergentSO5,CenkeAnders} We suggest that this continuous transition may be accessed by starting with the $S =1/2$ square lattice Neel-VBS transition and introducing a weak rectangular anisotropy (even weaker than considered in Ref.~\onlinecite{BlockMelko}). We also suggest that the same $O(4)$ symmetric CFT governs the Neel-VBS transition of the $S  =1$ square lattice.

We would like to note that some of our results have been recently independently obtained by other groups. Ref.~\onlinecite{CenkeAnom}  discusses  LSM like anomalies at deconfined critical points using less formal methods. Ref.~\onlinecite{ShinseiCryst} discusses LSM like anomalies in a number of gapless systems, including the 1+1D $S  =1/2$ chain. Ref.~\onlinecite{ZoharRyan}  provides a field-theoretic analysis of anomalies of the CP$^1$ model describing deconfined critical points in 1+1D and 2+1D - we give a slightly different derivation of these anomalies here and provide a physical interpretation.  While this manuscript was being completed Ref.~\onlinecite{ZoharWalls} appeared, which also discusses the implication of anomalies of 2+1D CP$^1$ model for lattice antiferromagnets.

This paper is organized as follows. In section \ref{sec:1d}, we discuss the anomalies of the 1+1D $S = 1/2$ chain: in \ref{sec:bos} we use the Abelian bosonization description of the chain, and in \ref{sec:CP1chain} - the CP$^1$ description. The latter allows for a more complete formal analysis where the $SO(3)_s$ and translational symmetries are gauged. Section \ref{sec:3Dpreamble} is devoted to the Neel-VBS deconfined critical point in 2+1D: the case of the $S  =1/2$ square lattice is discussed in \ref{sec:square}, and of the $S = 1/2$ honeycomb lattice in \ref{sec:honey}. A physical picture of the mixed anomaly involving the lattice rotational symmetry and $SO(3)_s$ is given in section \ref{sec:vort}: here we clarify the old arguments of Ref.~\cite{mlts04} regarding $S =1/2$ moment in the VBS vortex core. Section \ref{sec:S1} discusses some issues involving the breaking of continuous Lorentz (rotation) symmetry of the low-energy field theory description. Section \ref{sec:S1} also discusses anomalies of the $S =1$ deconfined critical point on the square lattice. Section \ref{sec:dyn} has a slightly different focus: it is devoted to the possibility that $S  =1/2$ rectangular lattice and $S =1$ square lattice Neel-VBS transitions might be continuous.  Concluding remarks are presented in section \ref{sec:future}. We also point out appendices \ref{app:CP1d}, \ref{app:CP2d}, which give a careful definition of the CP$^1$ model in 1+1D and 2+1D as a boundary of a higher dimensional SPT phase. Finally, appendix \ref{app:vort} discusses  VBS vortices in the context of the nearest neighbour dimer model, supplementing the discussion in section \ref{sec:vort}.





\section{$S = 1/2$ spin chain in 1+1D.}
\label{sec:1d}
We begin with the example of the $S = 1/2$ antiferromagnetic chain in 1+1D. While anomalies in this example have been studied at length before,\cite{ShinseiCryst,Oshikawa2017}  our interpretation of the ``emergent anomaly" and its consequences is somewhat different from that in the literature.

\subsection{Bosonized description}
\label{sec:bos}
We begin with the bosonized description of the chain (we work in real-time here),
\beq L = \frac{1}{2\pi} \d_t \theta \d_x \varphi - \frac{1}{4\pi} ((\d_x \varphi)^2 + (\d_x \theta)^2) \eeq
The microscopic operators are expressed as $S^+_j \sim A (-1)^{j} e^{i \theta}$, 
$S^z \sim A (-1)^j  \sin \varphi + \frac{1}{2\pi}  \d_x \varphi$, 
$V \sim \cos \varphi$, where $V_j \sim (-1)^{j} \vec{S}_j \cdot \vec{S}_{j+1}$ is the VBS order parameter. Here, we use Abelian bosonization, so only the $SO(2)_z$ subgroup of $SO(3)_s$ symmetry, corresponding to spin rotations around the $z$ axis, is manifest. (Below, we will also discuss the CP$^1$ formulation where the full $SO(3)_s$ symmetry is manifest). The $SO(2)_z$ symmetry acts as
\beq SO(2)_z:\,\, \theta \to \theta + \alpha, \quad \varphi \to \varphi\eeq
The translational symmetry acts as
\beq T_x:\,\, \theta \to \theta + \pi, \quad \varphi \to \varphi + \pi \label{Zxbos}\eeq
Note that the microscopic $Z$ translation symmetry acts in a $Z_2$ manner in the low-energy theory, so we will sometimes refer to $T_x$ as $Z^x_2$. 

Let's first discuss the manifestation of the LSM anomaly, which involves the combination of $SO(2)_z$ and translation symmetry $T_x$. First, consider a closed chain with an odd number of sites. Increasing the number of sites in the chain by one is tantamount to inserting a flux of the $T_x$ symmetry through the cycle of the chain. Using the action of $T_x$ (\ref{Zxbos}), a chain with an odd number of sites corresponds to twisted boundary conditions, $\theta(x + L) = \theta(x) + 2 \pi (n+1/2)$,\, $\varphi(x+L) = \varphi(x) + 2 \pi (m+1/2)$. Now, the total $SO(2)_z$ charge of the chain is $S^z = \frac{1}{2\pi} \int_0^L dx\, \d_x \varphi$. So we see that the chain with an odd number of sites carries $S^z$ which is half-odd-integer. Of course, this is precisely the correct physics for an $S  =1/2$ chain. However, if the microscopic symmetry was really $SO(3)_s$ (and its subgroup $SO(2)_z$) then only integer values of $S^z$ would be allowed - so our theory is anomalous.

Another (more standard) identification of the LSM anomaly proceed via threading flux of $SO(2)_z$ through the chain. When flux $\alpha$ of $SO(2)_z$ is threaded through the chain, the fields satisfy twisted boundary conditions, $\theta(x + L) = \theta(x) + 2\pi n+ \alpha $,\, $\varphi(x+L) = \varphi(x) + 2 \pi m$. Thus, as we insert flux $2\pi$ of $SO(2)_z$ the winding number of $\theta$ increases by $2\pi$, while the winding number of $\varphi$ remains unchanged. Now, from the action of translational symmetry (\ref{Zxbos}) we can identify the physical momentum
\beq P = \frac{1}{2} \int_0^L dx\, (\d_x \varphi - \d_x \theta)\eeq
So, after threading flux $2\pi$, the momentum $P$ changes by $\pi$. Of course, this is the result that we expect microscopically from the $S =1/2$ chain.\cite{LSM} However, if we treated $SO(2)_z$ and translation as on-site internal symmetries, then the momentum $P$ cannot change after flux-threading. So this again is a signature of the intrinsic LSM anomaly.

Next, we proceed to the emergent anomaly, which is associated  with the translation symmetry and does not require spin-rotations or time-reversal. We observe that the action of translational symmetry (\ref{Zxbos}) coincides precisely with the action of $Z_2$ symmetry on the edge of a 2+1D $Z_2$ protected SPT.\cite{ChenWen,LevinGu,luav12} The edge of a $Z_2$ protected SPT cannot be gapped out without breaking $Z_2$. Now, any translationally invariant weak perturbation that we add to the theory must respect $Z^x_2$, so such perturbations cannot open  a symmetric gap.

It is instructive to understand how the argument above breaks down when the perturbation added is not weak. Indeed, we know that, for instance, a sufficiently large uniform Zeeman field can fully polarize the spin chain. A weak Zeeman field corresponds to a perturbation,
\beq \delta L =  \frac{\delta}{2\pi} \d_x \varphi  \eeq
with $\delta \sim B_z$. This perturbation can be eliminated by redefining $\tilde{\varphi}(x) = \varphi(x) - \delta x$. Under translations by a lattice spacing $a$, $T_x: \, \tilde{\varphi}(x) \to \tilde{\varphi}(x + a) + \pi + \delta a$. Thus, translations no longer act on $\tilde{\varphi}$ in a $Z_2$ manner. As we keep increasing $B_z$, eventually we reach a point where, $T_x:\,\, \tilde{\varphi}(x) \to \tilde{\varphi}(x+a) + 2 \pi$, so a perturbation 
\beq \delta L \sim \cos \tilde{\varphi}  \label{Ltphi}\eeq
 becomes allowed and can open  a gap - this corresponds to a fully polarized chain. Physically, the momenta at which gapless degrees of freedom are present evolve as $B_z$ is tuned until momentum preserving backscattering terms are allowed. 
  If we express (\ref{Ltphi}) in terms of the original field $\varphi$, $\delta S \sim \int dx dt \cos (\varphi + \pi x/a)$. Clearly, close to the starting theory $\delta = 0$, this term vanishes since the momenta carried by the continuum field $e^{i \varphi}$ are assumed to be small (much smaller than $\pi$). 
  
The example considered here is quite general. Any continuum field theory where gapless degrees of freedom sit at isolated points in momentum space will have an emergent continuum translational symmetry (in our example, $\varphi(x) \to \varphi(x + \epsilon)$, $\theta(x) \to \theta(x + \epsilon)$). For ``kinematic" reasons outlined above, these continuum translations are preserved by any weak perturbation.  By combining these continuum translations with microscopic translations, we get a purely internal symmetry. If the underlying gapless excitations sit at commensurate points in momentum space, this internal symmetry will act as a finite group $G$ in the field theory ($Z_2$ in our example). $G$ might be an anomalous symmetry of the theory, in which case weak translation preserving perturbations cannot open  a gap. 

The example with the Zeeman field also illustrates how to immediately determine whether an anomaly is intrinsic (of LSM type) - i.e. whether it is stable to large perturbations away from a particular critical state. Again, for this purpose it suffices to treat translations as a purely internal symmetry, but one that acts in a $Z$ manner. To compute the anomaly, one can further restrict to Lorentz invariant theories (such as one describing the field $\tilde{\varphi}$ in the example above). For a $Z$ symmetry, the charge of the field can continuously change, e.g. $\tilde{\varphi} \to \tilde{\varphi} + \alpha$, with $\alpha$ -arbitrary, which in the example above ultimately removes the anomaly for translations.

We leave  the discussion of bulk-boundary correspondence for the anomalies described above to next section.


\subsection{CP$^1$ description.}
\label{sec:CP1chain}
We saw in the previous section that the 1+1D $S  =1/2$  spin chain possesses anomalies associated with the $SO(3)_s$ spin-rotation symmetry and translational symmetry. In this section, we discuss an interpretation of these anomalies when the chain is viewed as a surface of a 2+1D (crystalline) SPT phase.
Here we describe the gapless phase of the chain using the  CP$^1$ model  with a $\theta$ term at $\theta = \pi$,
\beq L = |(\d_{\mu} - i a_{\mu}) z_{\alpha}|^2 + i \theta  \frac{f}{2\pi}, \quad \theta = \pi \label{CP1d}\eeq
Here and below we work in Euclidean time. $a_{\mu}$ is a $u(1)$ gauge field and $f = \epsilon_{\mu \nu} \d_{\mu} a_{\nu}$ is the associated field strength. $z_\alpha$, $\alpha = 1, 2$, is a complex scalar transforming in the projective $S  =1/2$ representation of spin-rotation group $SO(3)_s$. The Neel order parameter is identified with $\vec{n} \sim z^{\dagger} \vec{\sigma} z$, and the VBS order parameter with $V \sim f$. 
Under translations by one lattice spacing 
\beq T_x: z\to i \sigma^y z^*, \,\, a \to - a\eeq
so that both $\vec{n}$ and $V$ are odd under $T_x$, as necessary. Note that $T^2_x z_\alpha = -z_\alpha$, i.e. $T^2_x$ is a rotation by $\pi$ in the $u(1)$ gauge group - i.e. $T^2_x$ acts trivially on all physical observables. This means that $T_x$ acts as a $Z_2$ symmetry in the field theory (\ref{CP1d}).  In a recent work \cite{ZoharRyan} it was shown that this $Z^x_2$ symmetry is anomalous. Moreover, the combination of $Z^x_2 \times SO(3)_s$ is also anomalous.\cite{ZoharDavide} In fact, as found in \cite{ZoharRyan}, one can think of (\ref{CP1d}) as living on the boundary of a $2+1$D SPT with $Z^x_2 \times SO(3)_s$ symmetry and bulk action,
\beq S_{bulk} = \pi i \int_{X_3} (x w^s_2 + x^3) \label{bulk3D}\eeq
where $X_3$ is the bulk three-manifold, $x \in H^1(X_3, Z_2)$ is the background gauge-field corresponding to $Z^x_2$ symmetry,  $w^s_2 \in H^2(X_3, Z_2)$ is the second Stiefel-Whitney class of the background $SO(3)_s$ bundle, and product of cohomology classes is the cup-product. We give a precise definition and a derivation of the bulk + boundary theory corresponding to (\ref{CP1d}), (\ref{bulk3D}) in appendix \ref{app:CP1d}. Note that our definition/derivation differs somewhat from the discussion in \cite{ZoharRyan}.   

We proceed to discuss the physical interpretation of the bulk action (\ref{bulk3D}). The first term in this action, 
\beq S_{1, bulk} = \pi i  \int_{X_3}  x w^s_2 \label{S1bulk}\eeq
 is precisely the intrinsic LSM anomaly for the combined $SO(3)_s$ and translational symmetry. The second term,
 \beq  S_{2, bulk} = \pi i \int_{X_3} x^3\eeq
  is the emergent anomaly for the translational symmetry alone. Let us begin with the emergent anomaly: we recognize that $S_{2,bulk}$ is precisely the bulk action of a $Z_2$ protected 2+1D SPT   in the presence of a background $Z_2$ gauge field $x$.\cite{DW,LevinGu} It is also immediately clear that this anomaly is not intrinsic if one remembers that the microscopic translation symmetry group is $Z$ rather than $Z_2$. The difference between a $Z$ gauge field and a $Z_2$ gauge field is that for a $Z$ gauge field $x$ (without vison defects) $d x = 0$, while for a $Z_2$ gauge field $d x  =0 \,(mod \,\,2)$ - the condition for a $Z$ gauge field is more restrictive.\footnote{Here and below, $d$ denotes the coboundary operation on cochains.} Now, for a $Z_2$ gauge field $x^2 = \frac{dx}{2} \, (mod \,\,2)$ (as cohomology elements). Therefore, if we interpret $x$ as a $Z$ gauge-field, $x^2 = \frac{dx}{2} = 0 \, (mod\,\, 2)$ and $S_{2, bulk}$ vanishes - no anomaly for translational symmetry alone is present. On the other hand, if we take $x$ to be a $Z_2$ gauge field then $S_{2, bulk}$ is generally non-vanishing.\footnote{As an example, consider $X_3 = \mathrm{RP}^3$ and $x$ -the generator of $H^1(\mathrm{RP}^3, Z_2)  =Z_2$.} As discussed in section \ref{sec:bos}, no translationally invariant weak perturbation of the critical chain breaks the internal $Z^x_2$ symmetry, therefore, to analyze the stability of the chain to weak perturbation one is allowed to couple it to a $Z_2$ gauge field, whereby one discovers an anomaly. To analyze stability to strong perturbations one must, however, treat $x$ as a $Z$ gauge field - then no anomaly is found and a gapped phase exists.

Next, we proceed to show that $S_{1,bulk}$ is the LSM anomaly. Here we may think of the bulk physically as a crystalline SPT obtained as a stack of Haldane chains  - the surface is then precisely an $S  =1/2$ chain.   $S_{1,bulk}$ is the ``response theory" of such a crystalline SPT. Let each Haldane chain stretch along the $y$ direction, and the chains be stacked along $x$. Let the length along $x$ be $L_x$ and the length along $y$ be $L_y$. For a moment, let both $x$ and $y$ be periodic, so that the space-time manifold is $S^1_x \times S^1_y \times S^1_\tau$. As noted in section \ref{sec:bos}, increasing $L_x \to L_x+1$ corresponds precisely to threading flux of the $T_x$ gauge field along the $x$ cycle. When $\int_{S^1_x} x  =1$ (and $x$ vanishes along the other cycles), $S_{1,bulk} = \pi i \int_{S^1_y \times S^1_\tau} w^s_2$ - which is precisely the response of the Haldane phase. Thus, as we increase $L_x$ by one, the system compactified along the $x$ direction goes from being a trivial $SO(3)_s$ SPT  to the  Haldane $SO(3)_s$ SPT. But that's precisely a property of a stack of Haldane chains! 

Another important manifestation of the LSM anomaly is obtained by thinking about the magnetic flux of $SO(3)_s$ in the 2+1D bulk. Let us compactify the bulk on $Y_2 \times S^1_\tau$, where we think of $Y_2$ as a spatial manifold. Place flux of $SO(3)_s$ through $Y_2$ (for instance, one can take $2\pi$ flux of the $SO(2)_z$ subgroup). The $SO(3)_s$ flux is defined only mod 2 and is measured precisely by $\int_{Y_2} w^s_2$. Therefore,  in this geometry, $S_{1,bulk} = \pi i \int_{S^1_\tau} x$. This means that an $SO(3)_s$ flux carries momentum $\pi$ under $x$ translations. But this is precisely right! Indeed, consider the bulk with a boundary. We may take the spatial bulk manifold to be a disc, so that the spatial boundary is a circle. Imagine moving the $SO(3)_s$ flux - e.g. $2\pi$ flux in the $SO(2)_z$ subgroup - from the trivial vacuum outside to inside the bulk. Outside the flux carries no momentum, but inside it carries momentum $\pi$. Therefore, in the process, there must be  momentum $\pi$ left on the boundary. That's precisely right! Indeed, from the boundary viewpoint, this process corresponds to threading $SO(2)_z$ flux $2\pi$ through the chain. We know microscopically that this changes the momentum by $\pi$. 

Note that the gauge fields $x$ that we considered in our discussion of $S_{1,bulk}$ satisfied $d x = 0$, i.e. the anomaly is already present when translations are treated as a $Z$ group. Again, this is what we expect for an intrinsic LSM anomaly. 

\section{Deconfined criticality in 2+1D}
\label{sec:3Dpreamble}
In this section we discuss the Neel to VBS transition in 2+1D on square and honeycomb lattices. The underlying field theory believed to  control this transition is the 2+1D CP$^1$ model,
\beq L = |(\d_{\mu} - i a_{\mu}) z_{\alpha} |^2  \label{eq:CP1xy} \eeq
where we use the same notation as in 1+1D, see section \ref{sec:CP1chain}. As written, the action (\ref{eq:CP1xy}) contains no monopole operators. Depending on the lattice and the value of spin $S$ one is considering, the action (\ref{eq:CP1xy}) admits various perturbations (particularly monopole operators) that we will discuss below. The continuum theory (\ref{eq:CP1xy}) has three internal global symmetries: i) $SO(3)_s$ rotations under which $z_{\alpha}$ transforms in the spinor representation. ii) $U(1)_{\Phi}$ flux symmetry under which the $2\pi$ flux monopole of $a$ that we denote by an operator $V$ transforms as 
\beq U(1)_{\Phi}: \,\, V \to e^{i \alpha} V\eeq
We denote the operator implementing a $U(1)_{\Phi}$ rotation by an angle $\alpha$ as $U^{\Phi}_\alpha$. 
iii) A {\it unitary} ``charge conjugation" symmetry: 
\beq C:\,\, z \to i \sigma^y z^*, \quad a \to -a, \quad V \to V^{\dagger} \eeq
Note that $C^2 = 1$ on gauge-invariant degrees of freedom, i.e. $C$ acts as a $Z_2$ symmetry. Combining $C$ with   $U(1)_{\Phi}$ we get a group $O(2)_{\Phi}$, therefore, the full internal symmetry group of (\ref{eq:CP1xy}) is $O(2)_{\Phi} \times SO(3)_s$. As we will discuss in the case of each lattice, the microscopic symmetries are implemented in the continuum theory as a subgroup of this  symmetry group (in the case of rotations, combined with continuum rotations). 

Before we specialize to particular physical symmetries, it is useful to compute the anomaly associated with the full continuum symmetry $O(2)_{\Phi} \times SO(3)_s$. This was done in Ref.~\onlinecite{ZoharRyan} (we give a slightly different derivation in appendix \ref{app:CP2d}). 
It was found that (\ref{eq:CP1xy}) is the boundary of a 3+1D $O(2)_{\Phi} \times SO(3)_s$ protected SPT with the following bulk response: 
\beq S_{bulk} = \pi i \int_{X_4} \, w_2[\xi_\Phi] \cup (w_2[\xi_s] + w^2_1[\xi_\Phi])\label{bulk4D}\eeq
Here, $X_4$ is the bulk four-manifold, $\xi_s$ is the $SO(3)_s$ bundle, $\xi_\Phi$ is the $O(2)_{\Phi}$ bundle, and as before $w_{1,2}$ denote the first and second Stiefel-Whitney classes. In particular, $w_1[\xi_\Phi] \in H^1(X_4, Z_2)$ is just the $Z_2$ gauge field corresponding to the charge-conjugation symmetry.

While this is not important for the anomaly analysis, let us say a few words about the order of transition in the continuum ``non-compact" theory (\ref{eq:CP1xy}). Numerical simulations suggest that it is either continuous or very weakly first order. Further, if the latter situation is the case, the weakly first order behavior is quasi-universal - the same critical exponents (and small drifts of these exponents with system size) are seen in microscopically different models.  Ref.~\onlinecite{DCPscalingviolations} proposed that this quasi-universal behavior  may be controlled by a nearby non-unitary critical point (or equivalently a unitary critical point appears if the parameters such as spatial dimension/number of flavors are varied slightly). Our discussion below can also be adapted to the quasi-universal first order scenario: in this case, when we talk of relevancy or irrelevancy of a certain operator in (\ref{eq:CP1xy}), we define it with respect to this non-unitary critical point/nearby unitary critical point. 

We now specialize to the particular lattices.

\begin{figure}[t]
\includegraphics[width=0.5\columnwidth]{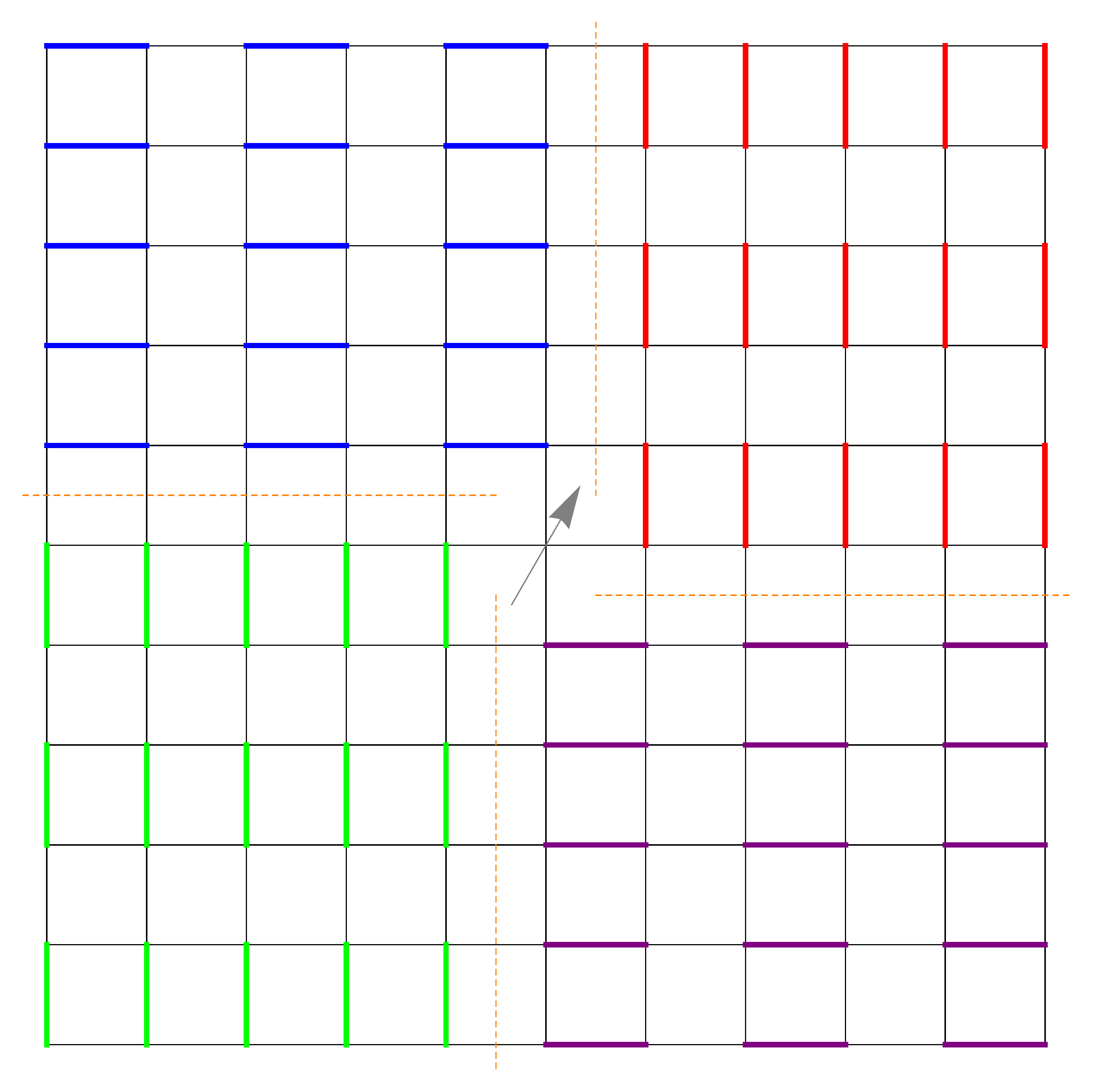}
\includegraphics[width=0.5\columnwidth]{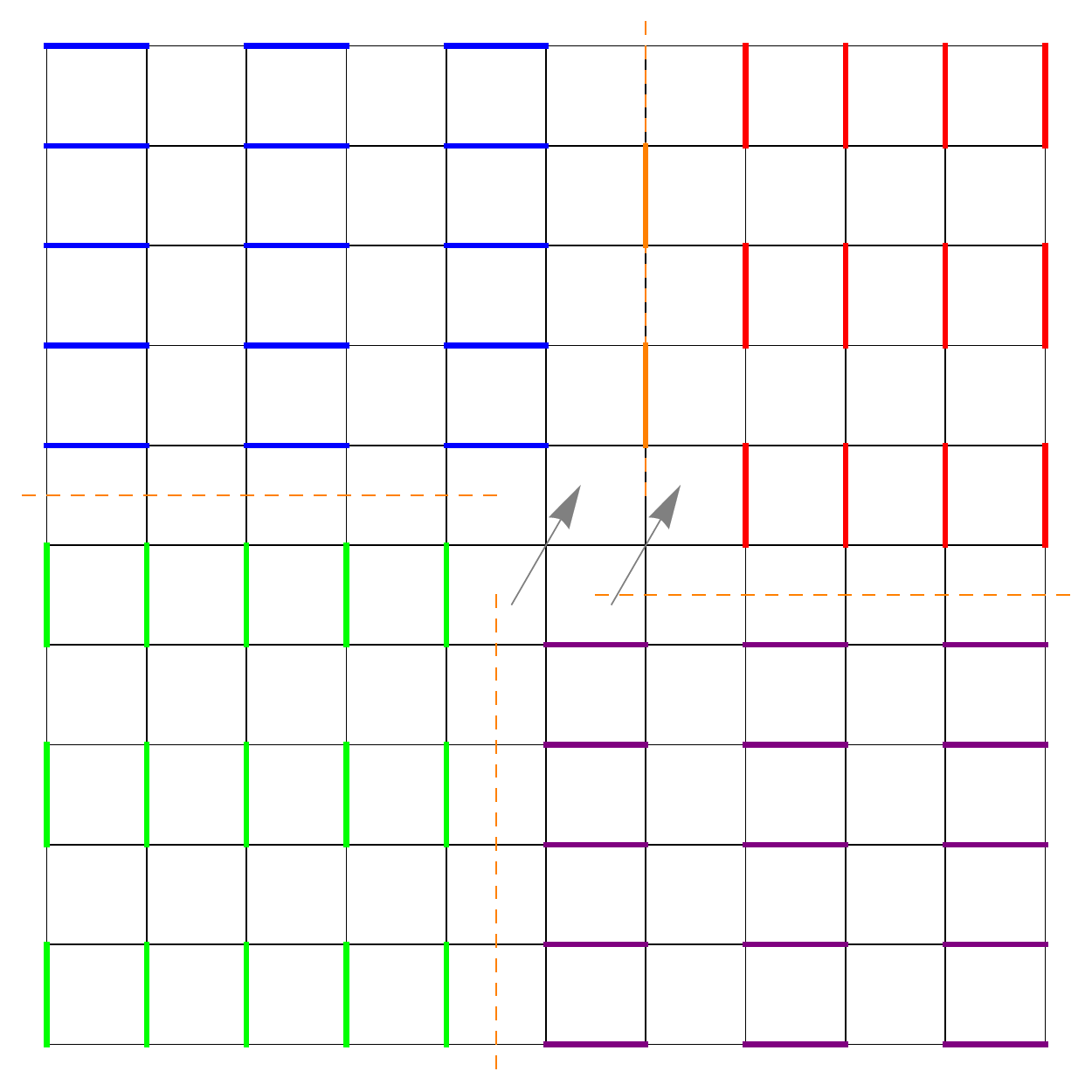}\\
\includegraphics[width=0.5\columnwidth]{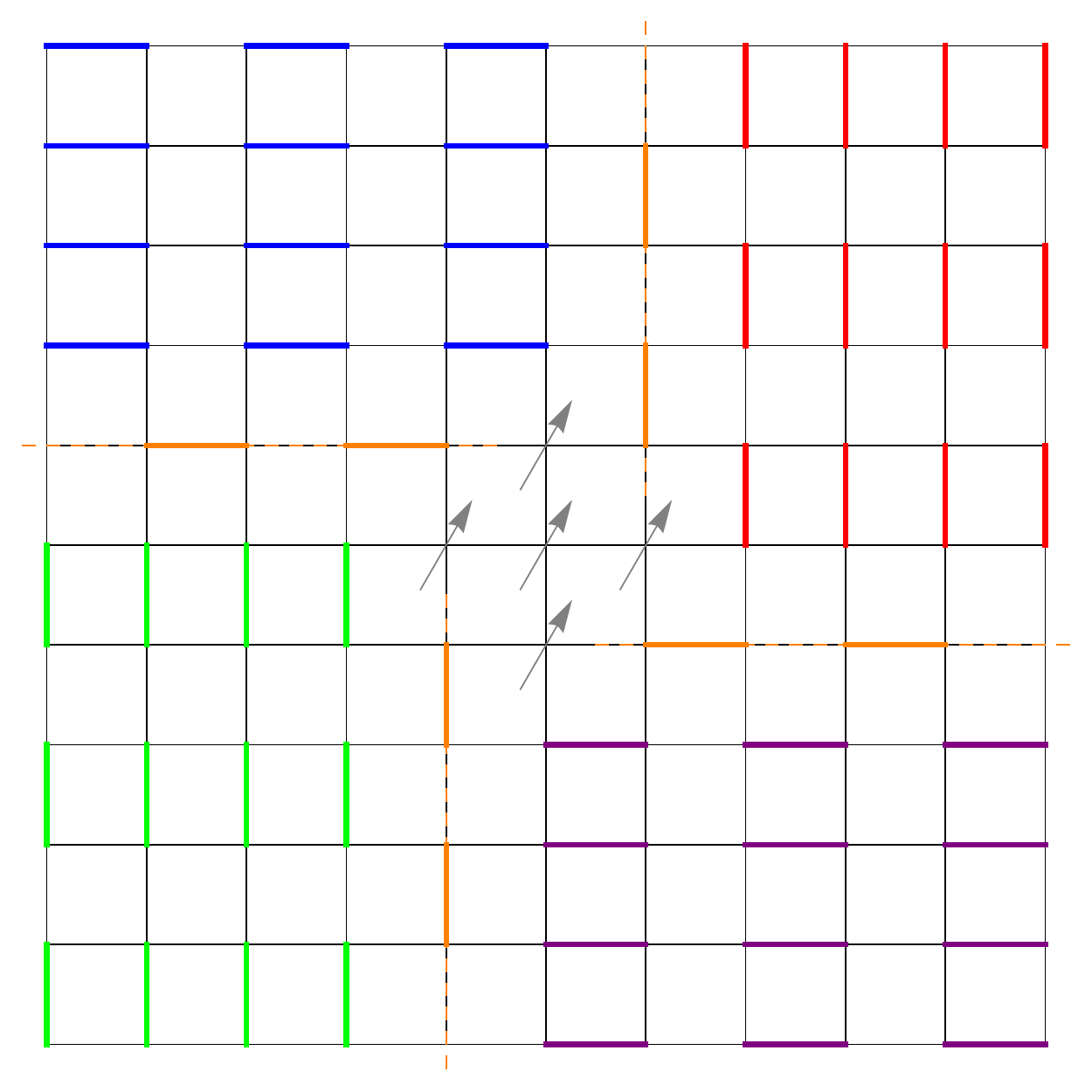}
\caption{Four domains of $S = 1/2$ square lattice VBS order with $V  =1, i, -1, -i$ in a $Z^{rot}_4$ vortex configuration. Domain walls are marked in dashed orange. Top left and bottom: a $Z^{rot}_4$ symmetric vortex traps half-odd-integer spin. Top right: a vortex which does not preserve the $Z^{rot}_4$ symmetry need not trap a spin (see also appendix \ref{app:vort}).}
\label{fig:vortsq1}
\end{figure}

\subsection{$S  =1/2$ square lattice}
\label{sec:square}

The lattice symmetries we focus on are translations $T_x$, $T_y$ and $Z_4$ rotations about a site. These act in the following way. The $Z_4$ rotation $R_{\pi/2}$ is just a $\pi/2$ rotation in the $U(1)_{\Phi}$ group $U^{\Phi}_{\pi/2}$ (together with a $\pi/2$ emergent continuum rotation), i.e.
\beq R_{\pi/2}: V(x) \to i V(R^{-1}_{\pi/2} x)\eeq
The translations act  $T_x = C$, $T_y =  U^{\Phi}_{\pi} C$, i.e. (apart from action on $z$, $a$),
\beq T_x: V \to V^{\dagger}, \quad T_y: V \to - V^{\dagger}\eeq
From these transformations we can identify $V = V_x + i V_y$ with the VBS order parameter (see Fig.~\ref{fig:vortsq1}). Further, we see that $T_x$, $T_y$ and $Z^{rot}_4$ act in the field theory as a $D_4$ subgroup of $O(2)_{\Phi}$, and the anomaly can be obtained by replacing the $O(2)_\Phi$ bundle $\xi_\Phi$ in (\ref{bulk4D}) by the $D_4$ bundle. Let us focus on two subgroups of this $D_4$. 

1). Imagine restricting the lattice symmetry to $Z_4$ rotations. Then we are interested in the $Z_4$ subgroup of $U(1)_{\Phi}$, so there are no $C$ gauge fields and $w_1[\xi_\Phi] = 0$. Further, for a $U(1)_{\Phi}$ gauge field, $w_2[\xi_\Phi] = \frac{F}{2\pi} \,(mod\,2)$, where $F \in H^2(X, Z)$ is the Chern class of the $U(1)_{\Phi}$ bundle. In our case, if we denote the $Z_4$ gauge field by $\gamma \in H^1(X_4, Z_4)$, $\frac{F}{2\pi} = \frac{d \gamma}{4} \in Z$. The anomaly (\ref{bulk4D}) then becomes
\beq S_{bulk} = \pi i \int_{X_4} \frac{d \gamma}{4} \cup w^s_2  \label{bulkrot} \eeq
This is a mixed anomaly involving  $Z_4$ rotations and $SO(3)_s$ symmetry. It is generally non-vanishing. Indeed, even if we restrict to only 180 degree  rotations, i.e. take $\gamma = 2 \tilde{\gamma}$ with $\tilde{\gamma} \in H^1(X_4, Z_2)$, the  action (\ref{bulkrot}) is still non-trivial,
\beq S_{bulk} = \pi i \int_{X_4} \frac{d \tilde{\gamma}}{2} \cup w^s_2 = \pi i \int_{X_4} \tilde{\gamma}^2 w^s_2  \label{bulkrot2} \eeq
The presence of the anomalies (\ref{bulkrot}), (\ref{bulkrot2}) is in exact accord with an LSM like theorem stating that a trivial gap is impossible  in a system with spin $S  =1/2$ located at a 180 degree rotation center.\cite{MikeHaruki} Thus, these anomalies are intrinsic anomalies.

2). Imagine restricting the lattice symmetry to translations $T_x$, $T_y$. In the field theory, these act as a $Z^x_2 \times Z^y_2$ subgroup of the $O(2)_\Phi$ group, corresponding to $O(2)$ transformations  ${\rm diag}(1,-1)$ and ${\rm diag}(-1,1)$. Denoting the $Z^x_2$ and $Z^y_2$ gauge fields as $x$ and $y$, we have $\xi_\Phi = \xi_{x} \oplus \xi_y$ - i.e. $\xi_\Phi$ is a direct sum of line bundles $\xi_x$ and $\xi_y$.  Using the Whitney formula, $w_1[\xi_\Phi] = w_1[\xi_x] + w_1[\xi_y] = x + y$, $w_2[\xi_\Phi] = w_1[\xi_x] w_1[\xi_y]  = x y$. So the anomaly reduces to 
\beq S_{bulk} = \pi i \int_{X_4} \left(x y  w^s_2 + x^3 y + x  y^3\right)   \eeq
The first term $x y w^s_2$ corresponds to the mixed LSM anomaly for the $SO(3)_s$ symmetry and translations. The last two terms comprise an emergent anomaly for the translation symmetry. 

Let's first discuss the LSM anomaly. Again, we can think of the $S  =1/2$ square lattice as a boundary of a stack of Haldane chains. We let the chains run along the $z$ direction and be stacked in a square lattice along $x$ and $y$. Let the $x$, $y$ and $z$ directions be periodic. Increasing $L_y$ by $1$ corresponds to threading $T_y$ flux along the $y$ cycle. Then, with such a $y$ flux, the bulk action becomes $S_{bulk} = \pi i \int_{S^1_x \times S^1_z \times S^1_\tau}  x w^s_2$. This is exactly the  action (\ref{S1bulk}) that we concluded corresponds to a 1d array of Haldane chains. This is the correct physics: fixing $L_y$ to be large by finite, we can view our bulk as a 2+1D phase protected by $T_x$ and $SO(3)_s$. At each $x$ ``site" there is a Haldane phase if $L_y$ is odd and an $SO(3)_s$ trivial  phase if $L_y$ is even.\footnote{We could have chosen a more general manifold $S^1_x \times Y_3$ with odd $x$ flux along $S^1_x$ to recover (\ref{bulk3D}). The choice of a three-torus for $Y_3$ is made for ease of  visualization and physical interpretation.} Again, we note that the action $S_{bulk} = \pi i \int_{X_4} x y  w^s_2$ is non-trivial even if $x$ and $y$ are $Z$ gauge-fields rather than $Z_2$ gauge fields, as befits an intrinsic LSM anomaly. 

We next discuss the emergent anomaly $S_{bulk} = \pi i \int_{X_4} \left(x^3 y + x  y^3\right)$. Again, if $x$ and $y$ are $Z$ gauge fields, then $x^2  = y^2 = 0 \,(mod \,\, 2)$ so $S_{bulk}$ vanishes. However, if $x$ and $y$ are $Z_2$ gauge-fields then $S_{bulk}$ is non-trivial - in fact, it is precisely the  response of a $Z_2 \times Z_2$ protected SPT in 3+1D.\footnote{Recall that  $Z^x_2 \times Z^y_2$ protected SPT phases in 3+1D are classified by $H^4(Z_2 \times Z_2) = Z^{(1)}_2 \times Z^{(2)}_2$. The generator $Z^{(1)}_2$ has the response, $S  =\pi i \int_{X_4} x^3 y$, and the generator $Z^{(2)}_2$, $S  =\pi i \int_{X_4} x y^3$. Our action is the sum of the two generators. Focus on one of the generators, $S  =\pi i \int_{X_4} x^3 y$. Consider $X_4 = S^1 \times Y_3$. Placing flux of $y$ through $S^1$ gives $S  =\pi i \int_{Y_3} x^3$ - the partition function of 2+1D $Z^x_2$ SPT on $Y_3$. Thus, threading flux of $Z^y_2$ through $S^1$ toggles between a trivial and non-trivial 2+1D $Z^x_2$ SPT. This is precisely the property of a $Z^x_2\times Z^y_2$ SPT in 3+1D.\cite{WenResponse,WangLevin1}} Since translations act in a $Z_2$ manner in the continuum field theory, we conclude that the $Z_2\times Z_2$ anomaly is relevant to the vicinity of the deconfined critical point. In particular, no weak translationally invariant perturbation can open a trivial gap (even if it breaks the spin-rotation symmetry).

Another emergent anomaly is present for the combination of diagonal translations $T_x T_y$ and $SO(3)_s$. In the continuum theory $T_x T_y$ acts in the same way as $180$ degree rotations, so, we indeed,  expect such a mixed anomaly. If only the $Z_2$ symmetry corresponding to $T_x T_y$ is gauged, we have $x = y$. Then 
\beq S_{bulk} = \pi i \int_{X_4} x^2 w^s_2 \eeq
which is again generally non-trivial for $x$ - a $Z_2$ gauge field, but vanishes for $x$ - a $Z$ gauge field. From a lattice perspective, we know that if we stagger the exchange strength as shown in Fig.~\ref{fig:stag}, for sufficiently strong staggering we will drive the system into a trivial gapped phase. However, the anomaly analysis above indicates that it does not occur for weak staggering.

\begin{figure}[t]
\includegraphics[width=0.7\columnwidth]{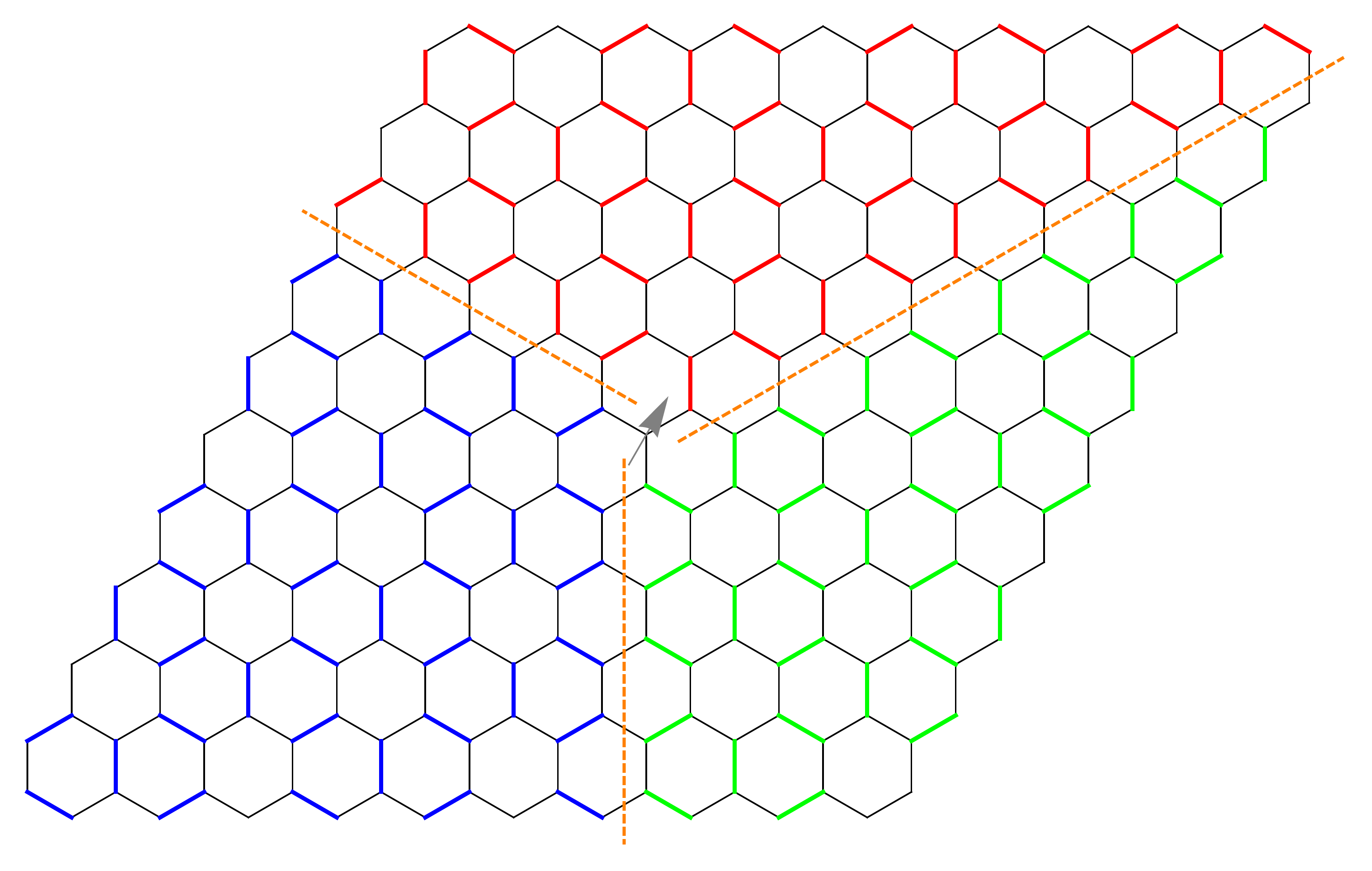}\\
\includegraphics[width=0.7\columnwidth]{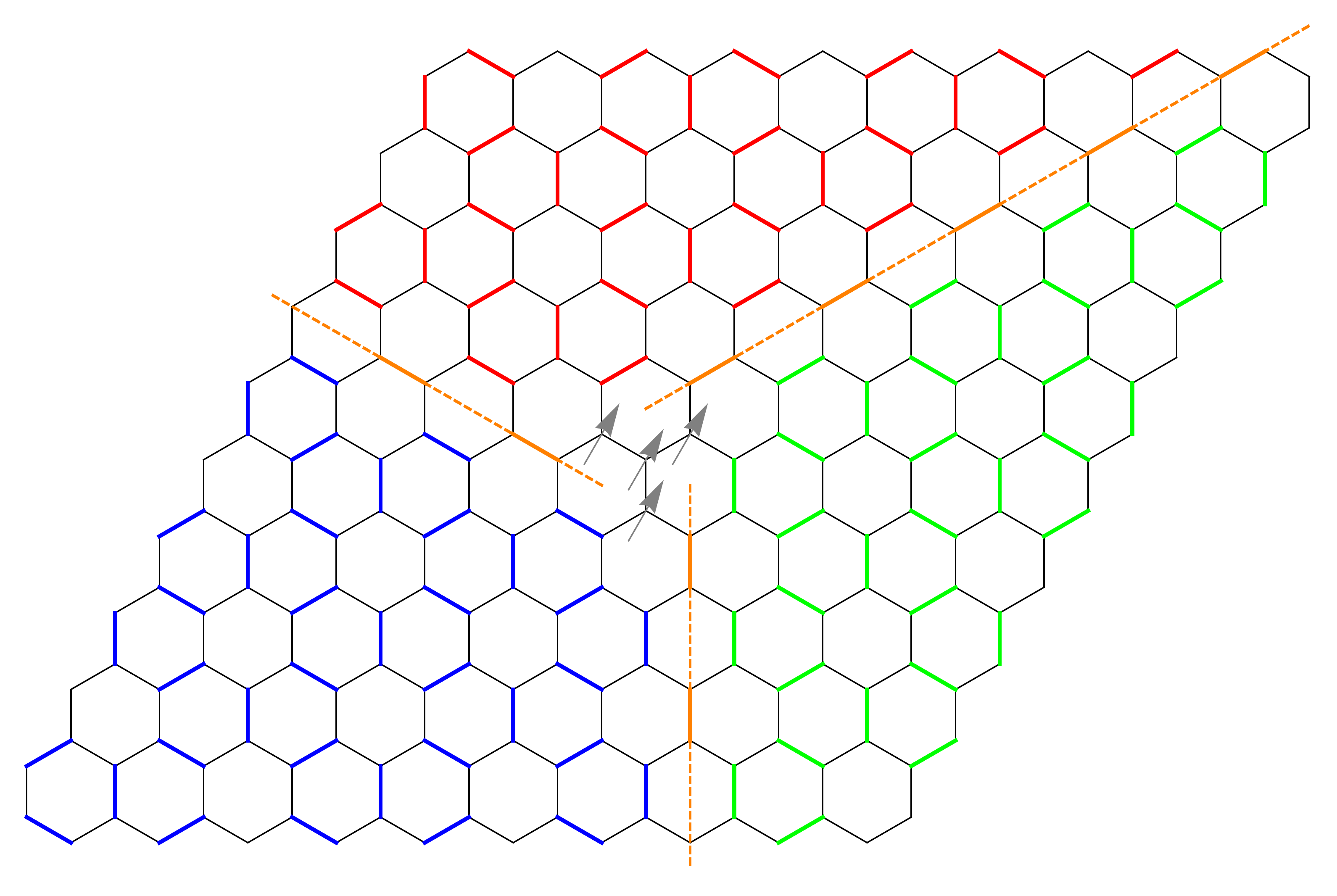}
\caption{Three domains of $S = 1/2$ honeycomb lattice VBS order with $V  =1, e^{2 \pi i/3}, e^{4 \pi i/3}$ in a $Z^{rot}_3$ vortex configuration. Domain walls are marked in dashed orange. A $Z^{rot}_3$ symmetric vortex may or may not trap $S  =1/2$ depending on the details of the domain walls.}
\label{fig:vorth}
\end{figure}

\subsection{$S =1/2$ honeycomb lattice}
\label{sec:honey}
We now discuss the case of the honeycomb lattice. The transition we consider is from a Neel phase to a Kekule-VBS phase (see Fig.~\ref{fig:vorth}). The symmetries we will be interested in are: $60$ degree rotations about a plaquette center $R_{\pi/3}$ and translations $T_1$, $T_2$ along the lattice vectors. These act as $T_1 = U^{\Phi}_{2 \pi/3}$, $T_2 = U^{\Phi}_{-2\pi /3}$, $R_{\pi/3} = C$, i.e.
\bea &&T_1:\,\,\,\,\,\, V\to e^{2 \pi i/3} V, \quad T_2: V \to  e^{-2\pi i/3} V \nn\\
&&R_{\pi/3}:\, V(x) \to V^{\dagger}(R^{-1}_{\pi/3}(x)) \eea
Thus, the monopole $V$ is identified with a Kekule-like VBS order parameter (see Fig.~\ref{fig:vorth}). Further, the lattice symmetries above act in the continuum theory as a $D_3$ subgroup of $O(2)_{\Phi}$. As discussed in appendix \ref{app:vanish}, for a $D_3$ bundle $w_2[\xi_\Phi] = 0$, so $S_{bulk} = 0$. Hence, in this case there are neither emergent nor intrinsic anomalies. The absence of an intrinsic anomaly is in agreement with the existence of a trivial gapped state on the honeycomb lattice.\cite{ZaletelHoney,YRanHoney} Let us now discuss possible consequences of the absence of emergent anomalies. The symmetries of the honeycomb lattice (in particular, the symmetries discussed above), permit a triple monopole perturbation,
\beq \delta L \sim V^3 + (V^{\dagger})^3\eeq
It is expected that this is the most relevant perturbation to the critical theory (\ref{eq:CP1xy}) describing the Neel to VBS transition on the honeycomb lattice (besides the perturbation $|z_\alpha|^2$ that tunes between the two phases). If this perturbation is irrelevant, the transition is described by the ``non-compact" theory (\ref{eq:CP1xy}) with an emergent $O(2)_\Phi$ symmetry, whose anomaly prohibits a trivial gap. On the other hand, if the perturbation is relevant then the symmetry of the low-energy theory is really only $D_3$. Since this symmetry is not anomalous, it is possible that a region of trivial gapped symmetric phase opens up between the Neel and VBS phases on the honeycomb lattice.\footnote{Strictly speaking, we also need to demonstrate that when the reflection/time-reversal symmetries are added to symmetries considered  above, no emergent anomalies are present. We leave this to future work.}

Numerically, the Neel-VBS transition on the honeycomb lattice appears continuous or very weakly first order.\cite{DamleH1, KawashimaH, BlockMelko, DamleH2} Further, on finite but large systems the same critical exponents are observed as on the square lattice. This suggests that the same ``non-compact" theory (\ref{eq:CP1xy}) governs the transition on the honeycomb lattice as on the square lattice. However, while on the square lattice nearly $U(1)$ symmetric histograms of the VBS order parameter are seen, which has been interpreted as evidence for the irrelevancy of the quadruple monopole operator $V^4$, on the honeycomb lattice a clear $Z_3$ anisotropy of the histogram is observed. Thus, it may be the case that the $V^3$ operator is close to marginality. If it is slightly relevant, it may be that system sizes where its effects start to play a role have not been reached in Refs.~\cite{DamleH1, KawashimaH, BlockMelko, DamleH2}. In this light, it would be interesting to numerically study the Neel-VBS transition for  the $S  =1/2$ honeycomb lattice in more detail. As already pointed out in Ref.~\cite{DamleH2}, it would be particularly interesting to look for new microscopic models realizing this transition with the hope that some of them have larger values of $V^3$ perturbation than those studied previously.

\subsection{Vortices and domain walls}
\label{sec:vort}
In this section, we give a more physical picture of the mixed anomaly between lattice rotational symmetry and $SO(3)_s$ clarifying the previous discussion in Ref.~\onlinecite{mlts04}. 

It has long been appreciated that the essential feature of the Neel-VBS transition on the square lattice is that VBS vortices carry spin $S = 1/2$.\cite{mlts04} At the field-theory level, this is seen as follows.\cite{SO5gang} Imagine first that no monopoles of $a$ are present in the action, so that the $Z^{rot}_4$ symmetry is dynamically enlarged to $U(1)_{VBS} = U(1)_{\Phi}$. To nucleate a vortex of $U(1)_{VBS}$ one couples the system to a background $U(1)_{\Phi}$ gauge-field $A$,
\beq L = |(\d_{\mu} - i a_{\mu}) z_{\alpha} |^2 + \frac{i}{2\pi} A \wedge da \eeq
and considers a configuration with flux $2 \pi$ of $A$. In order for this configuration to carry no $a$ charge (i.e. be gauge-invariant), we must additionally nucleate a $z_{\alpha}$ particle, so the vortex carries $S =1/2$. This matches the bulk anomaly (\ref{bulk4D}). Indeed, if we compactify the bulk theory (\ref{bulk4D}) on $S^2 \times Y_2$ with flux $2\pi$ of $U(1)_{\Phi}$ through $S^2$ then (\ref{bulk4D}) reduces to $S  =\pi i \int_{Y_2} w^s_2 $ - the partition function of a Haldane chain. Considering $Y_2$ to be open, we see that a monopole of $U(1)_{\Phi}$ is just like an end of a Haldane chain - i.e. it carries $S  =1/2$. When a monopole of $U(1)_{\Phi}$ sits in the 3+1D bulk, there is flux $2 \pi$ of $A$ eminating through  the 2+1D surface, so a VBS vortex is present on the surface and carries spin $1/2$.

Now, what happens when the $U(1)_{\Phi}$ symmetry is broken to $Z^{rot}_4$? If we work in the VBS phase, a VBS vortex will break up into a junction of four domain walls of $Z^{rot}_4$, see Fig. \ref{fig:vortsq1}. This vortex still traps $S = 1/2$ as is clear from Fig. \ref{fig:vortsq1} top, left. This is in agreement with the anomaly surviving when $U(1)_{\Phi} \to Z^{rot}_4$. A crucial point is that one must consider a vortex, which is invariant under $Z^{rot}_4$ (for an alternative viewpoint appropriate for the nearest neighbour dimer model, see appendix \ref{app:vort}). For instance, the configuration in Fig. \ref{fig:vortsq1} top, right has the same four domains as in Fig.~\ref{fig:vortsq1} top, left. However, it is not $Z^{rot}_4$ symmetric - one of the domain walls differs from the other three. We can think of this configuration as obtained from Fig. \ref{fig:vortsq1} top, left by dressing one of the domain walls with a Haldane chain. The Haldane chain contributes an extra $S = 1/2$ to the vortex, so that the total spin is an integer. If we, instead, dress all four domain walls with Haldane chains, so that the configuration is again $Z^{rot}_4$ symmetric, Fig. \ref{fig:vortsq1} bottom, we again have a half-odd-integer spin trapped in the vortex core. 

What about the $S =1/2$ honeycomb lattice? Here, the rotational symmetry of interest is $Z^{rot}_3$, corresponding to $2\pi/3$ rotations about a site.\footnote{This is a composition of $R^2_{\pi/3}$ - $2\pi/3$ rotation about a plaquette center and a translation by one lattice spacing $T_1$.} In the present case  there exist $Z^{rot}_3$ symmetric $Z^{rot}_3$ vortices with both half-odd-integer and integer spin - see Fig. \ref{fig:vorth}. Schematically, one goes from the $S =1/2$ vortex to an integer spin vortex by dressing each of the $Z_3$ domain walls with a Haldane chain. Indeed, in Fig.~\ref{fig:vorth} bottom, there are two $S  = 0$ states that the four ``dangling" $S  =1/2$'s can be locked into. These two states carry $Z^{rot}_3$ quantum numbers of $e^{\pm 2\pi i/3}$. This is not a projective representation of $Z^{rot}_3$ (in fact, there are no projective representations of $Z_3$) - it may be screened by local degrees of freedom to give a completely trivial vortex. This is consistent with the absence of an anomaly on a honeycomb lattice.  

\subsection{$S = 1$ square lattice and breaking of continuous rotation symmetry}
\label{sec:S1}

So far, when discussing the anomalies we've treated the translational symmetry and rotational symmetry as internal symmetries of the theory. More formally, the low energy theory (\ref{eq:CP1xy}) has a full emergent Poincare symmetry and we've combined elements of this Poincare symmetry with microscopic lattice symmetries to obtain purely internal symmetries. The anomalies associated with these internal symmetries allow us to place constraints on the dynamics when the Poincare symmetry is present. But what if it is broken? By comparing our anomaly computations so far with the microscopic LSM theorem, we guess that the anomaly found for the internal symmetry at the Lorentz invariant point is, in fact, the correct anomaly.

For instance, consider the case of $S =1/2$ square lattice. 
One allowed perturbation in this case is the quadruple monopole operator,
\beq \delta L \sim V^4 + (V^{\dagger})^4 \eeq
Throughout our discussion above, when we wrote $V^q$ - we understood this to be a Lorentz {\it scalar}, which creates $a$ flux $2 \pi q$. Such perturbations do not break the Lorentz symmetry, although they do break $U(1)_{\Phi} \to Z_4$. However, there also exist operators which carry quantum numbers under both the Lorentz symmetry and $U(1)_{\Phi}$; let us denote these by $V^{q}_{\ell}$, where $q$ is still the $U(1)_{\Phi}$ charge and $\ell$ is the angular momentum, such that under continuum spatial rotations $SO(2)_L$,
\beq SO(2)_{L}: \,\, V^{q}_{\ell}(x) \to e^{i \ell \alpha}V^q_{\ell} (R^{-1}_\alpha x)\eeq
(here, the subscript $L$ stands for Lorentz). Consider for instance the perturbation
\beq \delta L \sim V^{q = 1}_{\ell = -1} + V^{q =-1}_{\ell = 1} \label{l1} \eeq
While this  perturbation breaks $U(1)_{\Phi}$ and $SO(2)_L$ individually, it preserves their combination - i.e. the microscopic lattice rotation.  The microscopic LSM theorem for $Z^{rot}_4\times SO(3)_s$ symmetry\cite{MikeHaruki} tells us that such a perturbation (even if relevant) cannot open a trivial gap. Note that the perturbation (\ref{l1})  breaks the lattice translational symmetry. A perturbation consistent with all the symmetries of the square lattice is
\beq \delta L \sim V^{q = 2}_{\ell = 2} + V^{q = 2}_{\ell = -2} + V^{q = -2}_{\ell =2} + V^{q = -2}_{\ell = - 2} \label{eq:l2q2} \eeq
Again, LSM theorem guarantees that this perturbation cannot open  a trivial gap. In fact, this perturbation is very likely irrelevant: unitarity implies that the scaling dimension of an operator with angular momentum $\ell \neq 0$ satisfies $\Delta_{\ell} \ge l + D -2$, where $D$ is the space-time dimension\cite{RychkovReview}, so in our case, $\Delta_{\ell = 2} \ge 3$.\footnote{We thank Adam Nahum for pointing out this fact.} It is unlikely that an operator other than the energy-momentum tensor exactly saturates the unitarity bound (if it does, it gives rise to a conserved $\ell  =2$ current). The numerically observed emergent $U(1)_{VBS}$ symmetry of the deconfined critical point\cite{SandvikJQ,emergentSO5} is also consistent with the irrelevancy of (\ref{eq:l2q2}).

\begin{figure}[t]
\includegraphics[width=0.5\columnwidth]{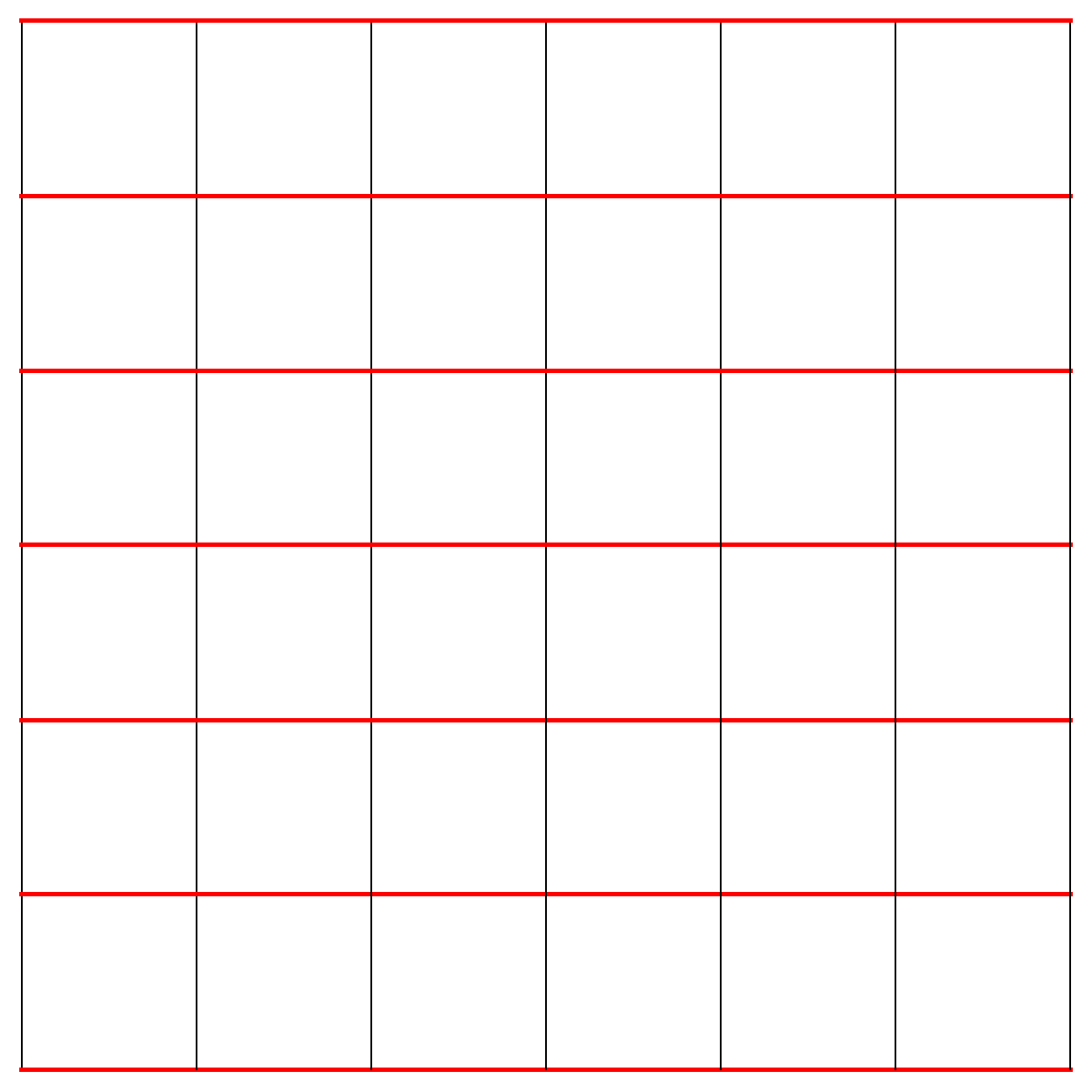}
\includegraphics[width=0.5\columnwidth]{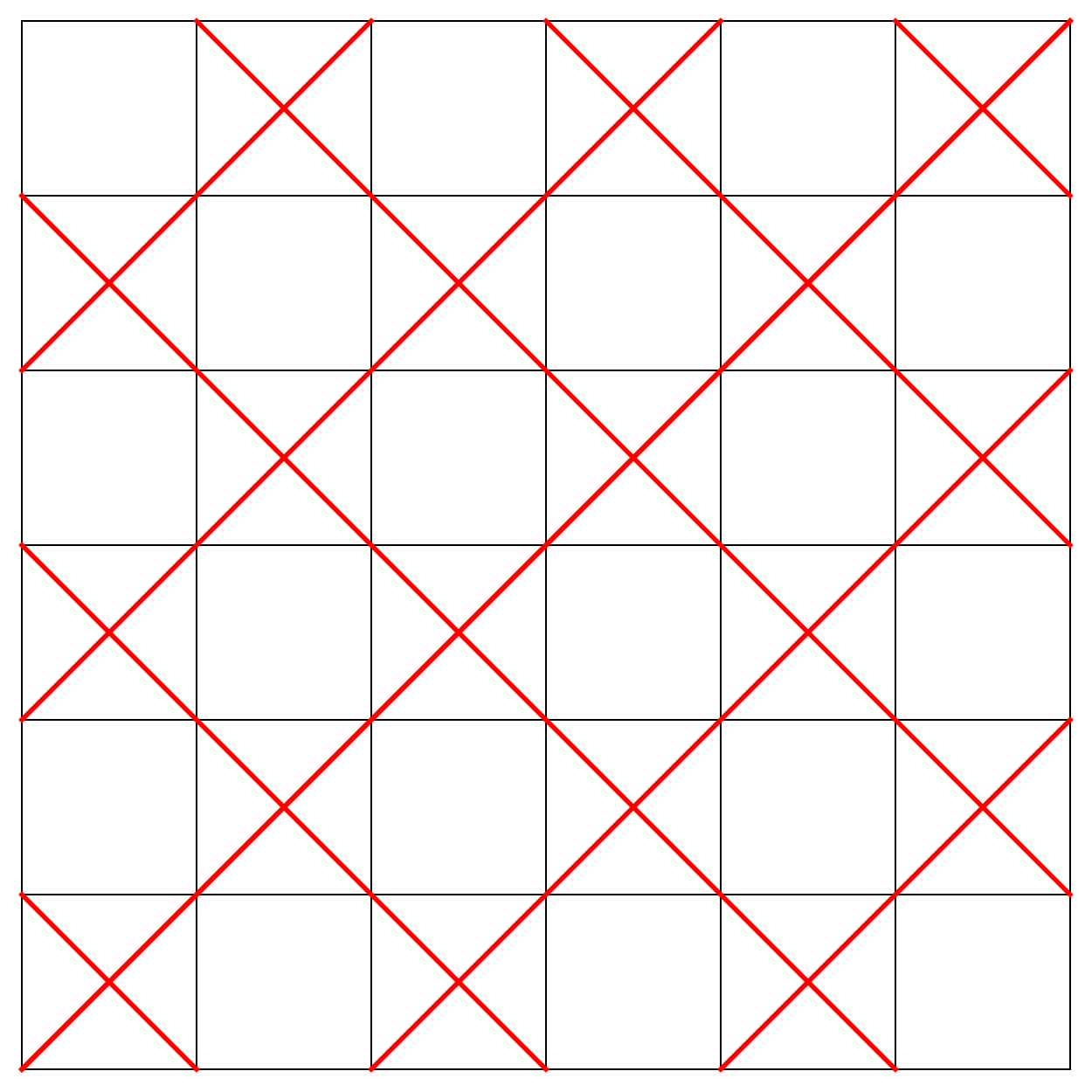}
\caption{$S =1$ square lattice VBS configurations. Red lines correspond to $S = 1$ spins locked into Haldane chains. Left: $Re(V) > 0, Im(V) = 0$. Right: $Re(V) = 0, Im(V) > 0$. Note that $Re(V)$ and $Im(V)$ are not related by any symmetry.}
\label{fig:S1}
\end{figure}

With the above remarks in mind, we proceed to the case of
 $S = 1$ Neel-VBS transition on the square lattice (see Fig.~\ref{fig:S1}). Here  the  symmetries are implemented in the following way: $T_x = C$, $T_y = C$, $R_{\pi/2} = U^{\Phi}_{\pi}$, i.e.
\bea &&T_x:\,\, V \to V^{\dagger}, \quad T_y: \,\, V \to V^{\dagger}\nn\\
&&R_{{\pi}/{2}}:\,\, V(x) \to - V(R^{-1}_{{\pi}/{2}} x)\eea
Note that when combined with the spacial rotation in the Lorentz group the 90 degree rotation symmetry acts in a $Z_2$ manner.  So when treated as internal symmetries, $T_x, T_y, R_{\pi/2}$ act as a $Z_2 \times Z_2$ subgroup of $O(2)_{\Phi}$. Since $T_x$ and $T_y$ act in the same way, let's focus on just one of them, say $T_x$. Denote the $Z_2$ gauge field corresponding to $T_x$ as $x$, and the $Z_2$ gauge field corresponding to $R_{\pi/2}$ as $\alpha$. The $O(2)_{\Phi}$ bundle $\xi_{\Phi}$ is then a direct sum of two $Z_2$ bundles: $\alpha$ and $x+\alpha$, so $w_1[\xi_{\Phi}] = x$, $w_2[\xi_{\Phi}] = \alpha (x+\alpha)$. The bulk action (\ref{bulk4D}) then is
\beq S_{bulk} = \pi i \int_{X_4} \left( (\alpha x + \alpha^2) w^{s}_2 + \alpha x^3\right) \label{S1anom}\eeq
which is generally non-vanishing for arbitrary $Z_2$ gauge fields $x$, $\alpha$. This anomaly implies that as long as we allow only Lorentz invariant (more specifically rotationally invariant) perturbations to the action (\ref{eq:CP1xy}) no trivial gap is possible. However, we know from an explicit construction that a trivial gapped state does exist for an $S  =1$ square lattice.\cite{ZaletelHoney} So there  must be no intrinsic anomaly present. To see this, we note that the microscopic symmetry group generated by $T_x, T_y$ and $R_{\pi/2}$ is actually $(Z^x \times Z^y) \rtimes Z^{rot}_4$.  As shown in appendix \ref{app:vanish}, for an $O(2)_{\Phi}$ bundle corresponding to this group, $w_2[\xi_{\Phi}] = 0$, so $S_{bulk} = 0$ in accord with microscopics.

This leaves the question: if we allow for weak Lorentz breaking perturbations to the CP$^1$ model consistent with $S = 1$ square lattice symmetry, can a trivial gap be opened?\footnote{This question is quite formal since in a microscopic lattice model there is no way to control the strength of Lorentz breaking perturbations.} For instance, we can envision a perturbation:
\beq \delta L \sim V_{\ell = 2}^{q = 1} + V_{\ell= -2}^{q=1} + V_{\ell = 2}^{q = -1} + V_{\ell = -2}^{q = -1} \label{S1l2}\eeq
which preserves both translation and rotation symmetry. Again, unitarity implies that the scaling dimension of this operator is greater or equal to $3$, so it is likely irrelevant. Suppose we did not know this fact, or wish to consider the combined effect of this perturbation and other relevant perturbations. It turns out that just from anomaly considerations, we can argue that (\ref{S1l2}) cannot open a trivial gap. Indeed, $T_x$ and $R_{\pi/2}$ act in the continuum theory as a $Z^x_2 \times Z^{rot}_4$ symmetry. Note that since the action no longer possesses Lorentz symmetry, rotations must be treated as a $Z_4$ group rather than $Z_2$ group. On the other hand, for weak perturbations, we may still continue to treat $T_x$ as a $Z_2$ symmetry. Then $\alpha^2 w^s_2 = \frac{d \alpha}{2} w^s_2 = 0 \,\,({\rm mod}\,\,2)$, since $d\alpha = 0\,\, ({\rm mod}\,\,4)$. However, the other two terms in (\ref{S1anom}) are generally non-trivial:
\beq S_{bulk} \to  \pi i \int_{X_4} \left( \alpha x  w^{s}_2 + \alpha x^3\right) \eeq
so a trivial gap cannot be opened.




\section{$S = 1/2$ rectangular lattice and $S  =1$ square lattice: dynamics}
\label{sec:dyn}
The present section has a slightly different emphasis from the rest of the paper. Here, we discuss a possibility that the Neel-VBS transition of $S  =1/2$ rectangular lattice and $S = 1$ square lattice can be continuous and described by a CFT with an emergent $O(4)$ symmetry. The same CFT  has been proposed to describe the $S =1/2$ easy-plane Neel-VBS transition on a square lattice (see Ref.~\onlinecite{SO5gang} and references therein).

Let us begin with the case of $S  =1/2$ rectangular lattice. To obtain the critical theory, we may start with the square lattice and weakly break the 90 degree rotation symmetry to the 180 degree rotation symmetry. One  perturbation to (\ref{eq:CP1xy}) this induces is
\beq \delta L = -\lambda_2 (V^2 + (V^{\dagger})^2) \label{V2} \eeq
Numerical simulations indicate that this operator is relevant.\cite{emergentSO5} However, this does not necessarily imply that it drives the transition first order. Recall that numerics suggests that the theory (\ref{eq:CP1xy}) possesses an emergent $SO(5)$ symmetry, with the Neel and VBS order parameters forming an $SO(5)$ vector $\vec{X} = (n^x, n^y, n^z, V^x, V^y)$. We can also form a traceless symmetric $SO(5)$ tensor, $X_{ab}$, $a, b = 1\ldots 5$, which is schematically $X_{ab} = X_a X_b - \frac{\delta_{ab}}{5} \vec{X}^2$. The operator $V^2$ is identified with $V^2 \sim X_{44} - X_{55} + 2 i X_{45}$. On the other hand, the operator $|z|^2$ which drives the phase transition on the square lattice is,  $|z|^2 \sim X_{44} + X_{55}$. So, on a rectangular lattice the $SO(5)$ invariant CFT is perturbed by
\beq \delta L = -\lambda_1 (X_{44} + X_{55}) - \lambda_2 (X_{44} - X_{55})\eeq
Crucially, the perturbations $\lambda_1$ and $\lambda_2$ are part of the same $SO(5)$ multiplet!\cite{emergentSO5} Now, without loss of generality, assume $\lambda_2 > 0$. If we tune the system to the point $\lambda_1 = -\lambda_2$, we have
\beq   \delta L = 2 \lambda_2 X_{55} \label{X55} \eeq
i.e. the system possesses an emergent $SO(4)$ symmetry at this point. In fact, this is the same perturbation of the $SO(5)$ invariant CFT that describes the easy-plane $S = 1/2$ deconfined critical point on the square lattice. In the CP$^1$ language the easy-plane deformation is simply an anisotropy,
\beq \delta L \sim \lambda_3 \left[ (|z_1|^2 - |z_2|^2)^2- \frac{1}{3} |z|^4\right] \eeq with $\lambda_3 > 0$. In the $SO(5)$ language this translates to,
\beq \delta L = -\lambda_1 (X_{44} + X_{55})+  \lambda_3 X_{33}\eeq
The transition point is now $\lambda_1 = 0$, which has exactly the same form as (\ref{X55}) (up to an $SO(5)$ rotation exchanging $X_3$ and $X_5$). 

Previously, it was thought that the easy-plane transition is first-order. However, recent simulations\cite{CenkeAnders} suggest that when the easy-plane anisotropy $\lambda_3$ is small, the transition is actually continuous and described by an $O(4)$ invariant CFT where the $O(4)$ vector is $\vec{Y} = (n^x, n^y, V^x, V^y)$.\footnote{The numerical evidence for the emergent $SO(4)$ symmetry comes from the fact that the critical exponents of the easy-plane Neel-VBS transition match with those of a different model with an explicit $SO(4)$ symmetry. The latter model realizes a transition between a trivial insulator and a bosonic integer quantum Hall state.\cite{CenkeAnders}} The transition on the rectangular lattice is then described by the same $O(4)$ invariant CFT with the $O(4)$ vector $\vec{Z} = (n^x, n^y, n^z, V^x)$. (As already noted, this possibility was first raised some time ago in Ref.~\onlinecite{tsmpaf06}). If we form the $SO(4)$ traceless symmetric tensor, $Z_{ab}$, then the perturbation driving the Neel-VBS transition on the rectangular lattice is
\beq \delta L \sim Z_{44} \label{Z44}\eeq
which breaks the emergent $O(4)$ symmetry to $SO(3)_s\times Z^{rot}_2 \times Z^x_2$. 
This should be compared to the perturbation driving the easy-plane square lattice transition
\beq \delta L \sim Y_{33} + Y_{44}\eeq
The perturbations driving the transitions in the two cases are different (albeit in the same multiplet), so the phases are also different (e.g. the Neel phase in the easy-plane case has only one Goldstone, while it has two Goldstones in the $SO(3)_s$ case). As for other perturbations on the rectangular lattice besides (\ref{Z44}), we have e.g. the component of a four-index traceless symmetric tensor $Z_{4444}$. This should be compared to a perturbation of the easy-plane theory $\sum_{a =1}^{2}\sum_{b= 3}^{4} Y_{aabb}$, which is in the same multiplet. This perturbation must be irrelevant for the easy-plane transition to be continuous and to possess and $SO(4)$ symmetry (as numerics suggest).

So far, we've only discussed Lorentz invariant perturbations on the rectangular lattice. There are also symmetry allowed Lorentz breaking perturbations. The most simple of these is $|D_x z|^2 - |D_y z|^2$, which however, can be eliminated by a coordinate rescaling. We assume that other Lorentz breaking perturbations are irrelevant.

For the case of $S  = 1$ magnet on a square lattice the double monopole perturbation (\ref{V2}) is again allowed, so we again conjecture a transition described by the same $O(4)$ invariant CFT.  Note that a set of Lorentz breaking perturbations distinct from those of a rectangular lattice are allowed here, e.g. Eq. (\ref{S1l2}). We again assume that these perturbations are irrelevant. 

\section{Future directions}
\label{sec:future}
In this paper, we have focused on the anomalies of lattice systems  associated with the combination of  spin-rotation symmetry and lattice translations/rotations. It will be interesting to extend this analysis to include time-reversal and reflection symmetries. In particular, it will be interesting to see if there are any emergent anomalies associated with these symmetries in the vicinity of the deconfined QCP on the honeycomb lattice (we expect that there is no intrinsic anomaly, since a trivial symmetric gapped state on the honeycomb lattice exists). If no emergent anomaly is found then an intermediate trivial phase whose appearance is driven by the $V^3$ operator might, indeed, be possible.  

The entire anomaly analysis carried out in this paper has been performed by tuning the critical theory to a Lorentz invariant point and treating lattice symmetries as internal symmetries. While our results agree with LSM-like theorems, this procedure is still very much a conjecture. A stronger argument in favor of this conjecture (perhaps, utilizing the technology of Ref.~\onlinecite{RyanDom}) is left to future work. 

Finally, in this work we have not considered LSM-like theorems relying on (usually fractional) $U(1)$-number filling per unit cell. Additional  subtleties arise in the formal treatment of this situation, so we leave it to future investigation.

\acknowledgements
We are grateful to M.~Cheng, D.~Else, I.~Kimchi, A.~Nahum, T.~Senthil, A.~Vishwanath for helpful discussion. R.~T. is supported by an NSF GRFP
grant.

\appendix
\section{CP$^1$ model in $1+1$D}
\label{app:CP1d}
In this appendix we deduce the bulk action (\ref{bulk3D}), which matches the anomalies of the 1+1D CP$^1$ model at $\theta = \pi$,
\beq L = |(\d_{\mu} - i a_{\mu}) z_{\alpha}|^2 + i \theta  \frac{f}{2\pi}, \quad \theta = \pi \label{CP1dapp}\eeq
 Let us begin by considering just the $Z^x_2$ symmetry and ignore $SO(3)_s$. Let us attempt to gauge the global $Z^x_2$ symmetry of (\ref{CP1dapp}). Then the scalar $z$ sees a combination of transition functions in the $u(1)_g$ gauge group and in the $Z^x_2$ group. Since, $T^2_x = u^g_{\pi}$, overall $z$ sees transition functions in $pin(2)_-$. 

Now, the immediate difficulty that one is faced with when trying to gauge $Z^x_2$ symmetry is how to define the $\theta$ term in (\ref{CP1dapp}). Indeed, locally $f \to -f$ under $Z^x_2$, so as written, the $\theta$ term is not well-defined. Instead, when $Z^x_2$ is gauged, we will define the theory in the following way. We think of the theory as living on the surface of a 2+1D SPT for the $Z^x_2$ symmetry. We call the bulk three manifold $X_3$ and the surface $M = \d X_3$. There is a $Z^x_2$ gauge field $x \in H^1(X_3, Z_2)$ living in the bulk and on the surface. On the surface, $x$ together with the $u(1)_g$ gauge field $a$ form a $pin(2)_-$ gauge field (note, $a$ lives only on the boundary $M$, {\it not} in the bulk $X_3$). Let's call the $pin(2)_-$ gauge bundle $\xi_g$. We find a three manifold $Y_3$ such that $\d Y_3 = M$ and $\xi_g$ extends to $Y_3$ as a $pin(2)_-$ bundle (therefore, $x$ also extends to $Y_3$).   We define the action of our theory as
\beq S_{bulk + bound} = S_{bound}[M] + \pi i \int_{X_3 \cup \bar{Y}_3} x^3 \label{bulkbound3D1} \eeq
with
\beq S_{bound}[M] = \int_M d^2 x \sqrt{g}\, (\d_{\mu} + i a_{\mu})z^{\dagger}   (\d^{\mu} - i a^{\mu})z  \label{bound1dr} \eeq
Note, $Y_3$ is not the ``physical" bulk manifold $X_3$ but rather an auxiliary manifold used to define the action. 
Further observe that the ``boundary" action (\ref{bound1dr}) is purely real and contains no topological terms. All the topological terms have been shifted to the second term on the RHS of  (\ref{bulkbound3D1}). While it is not immediately obvious, we will shortly show that when the $Z^x_2$ gauge field on the physical space $X_3$ is absent, (\ref{bulkbound3D1}) reduces to our original theory (\ref{CP1dapp}).

In order for (\ref{bulkbound3D1}) to be a well-defined action on a ``physical" bulk $X_3$ with a boundary $M$, we have to make sure that it is independent of the manifold $Y_3$ and the particular extension of the boundary $pin(2)_-$ bundle to $Y_3$ that we have chosen. To see this, it suffices to show that for a $pin(2)_-$ gauge field on a closed manifold $Y_3$, $\int_{Y_3} x^3 = 0 \,\, (mod\, 2)$. Indeed, if we project our $pin(2)_-$ bundle $\xi_g$ to an $O(2)$ bundle $\tilde{\xi}_g$, $x = w_1[\tilde{\xi}_g]$. Further, an $O(2)$ bundle has a lift to $pin(2)_-$ if and only if $w_2[\tilde{\xi}_g] + w^2_1[\tilde{\xi}_g] = 0$.\cite{KirbyTaylor} Thus, $w_2[\tilde{\xi}_g] = x^2 = \frac{dx}{2}$. Furthermore, $w_3 = w_1 w_2 + \frac{d w_2}{2}$. For an $O(2)$ bundle, $w_3 = 0$, so $w_1 w_2 = \frac{d w_2}{2} = 0$, i.e. $x^3  = 0$ and $\int_{Y_3} x^3 = 0$ for $Y_3$ - closed (note, $x^3  =0$ and prior relations hold only in the sense of $Z_2$ cocycles, so it is important for $Y_3$ to be closed! In particular, we cannot just drop the $Y_3$ part of (\ref{bulkbound3D}) - in fact, the resulting expression will not be gauge invariant).

We note that while (\ref{bulkbound3D1}) does not depend on $Y_3$, it clearly depends on the gauge field $x$ on the ``physical" three dimensional manifold $X_3$.  Crucially, the boundary $pin(2)_-$ bundle need not extend to the ``physical" bulk $X_3$, so in general $\int_{X \cup \bar{Y}}  x^3 \neq 0$. Indeed, when $X_3$ has no boundary, (\ref{bulkbound3D1}) reduces to (\ref{bulk3D}), which is the topological response of a $Z_2$ protected SPT. This tells us that the surface theory has a $Z^x_2$ anomaly.

It remains to show that (\ref{bulkbound3D1}) coincides with (\ref{CP1dapp}) when the $Z^x_2$ symmetry on $X_3$ is not gauged, i.e. when $x = 0$ on $X_3$. The boundary $M$ of $X_3$ is an oriented surface with a $u(1)$ gauge field $a$. When the flux $m = \int_M \frac{f}{2\pi} \in Z$ is not zero, we cannot extend $a$ from $M$ to some $Y_3$ as a $u(1)$ gauge field. However, as we will now show, we can extend $a$ from $M$ to $Y_3$ as a $pin(2)_-$ gauge field. First, it suffices to consider the case when $M$ is a two-sphere $S^2$ with flux $2\pi$.  Indeed, $M$ is always bordant to $m$ such spheres. So specializing to $M$ - a two-sphere $S^2$ with flux $2 \pi$, we must show that $\int_{X_3 \cup \bar{Y}_3} x^3 = 1$, so that the topological part of the action is given by $\pi i$, in accord with (\ref{CP1dapp}). We take $Y_3$ to be RP$^3 \setminus D_3$, where $D_3$ is a 3-dimensional ball. It is convenient to think of RP$^3$ as a three-dimensional ball of radius $R$ with antipodal points on the boundary identified. We obtain $Y_3$ by cutting out a ball of radius $1$ centred at the origin  from this realization of RP$^3$ (we take $R > 1$). The boundary  $M$ of $Y_3$ is  a sphere $S^2$ of radius $1$.  We place flux $2 \pi$ on this sphere. In polar coordinates, we choose 
\beq a_{\varphi}(r, \theta, \varphi) = \frac{1}{2} (1-\cos \theta), \quad a_\theta = 0, \quad a_r  =0. \eeq
Now, we must glue the fields at $r = R$. Clearly, we need to use the $Z^x_2$ symmetry to do so. We impose at $r  =R$,
\beq z(x) = e^{i \alpha(x)} i \sigma^y z^*(\iota(x)), \quad a(x) = - (\iota^* a)(x)+ d \alpha(x) \eeq
where $\iota: \theta \to \pi -\theta, \,\, \varphi \to \varphi + \pi$ is the antipodal map and $e^{i \alpha(x)}$ is a $u(1)$ gauge rotation. Choosing $e^{i \alpha(x)} = e^{i \varphi}$ does the job, leading to a consistent gluing condition. Thus, we have succeeded in extending the $pin(2)_-$ bundle to $Y_3$. The corresponding $Z^x_2$ gauge field $x$ on $Y_3$ integrates to $1$ along any loop connecting the antipodal points of the sphere $r = R$.  It remains to evaluate the topological action $\int_{X_3 \cup \bar{Y}_3} x^3$. Since $x = 0$ on $X_3$ we might as well replace $X_3$ by a ball of radius $1$ so that $X_3 \cup Y_3$ is RP$^3$. Clearly, $x$ is just the generator of $H^1(\mathrm{RP}^3, Z_2)$ so $\int_{X_3 \cup \bar{Y}_3} x^3 = 1$. QED.

So far, we have only attempted to gauge the $Z^x_2$ symmetry. Now, we will in addition gauge the $SO(3)_s$ symmetry. Again, we think of the system as living on the boundary of a 3D SPT with both $Z^x_2$ and $SO(3)_s$ symmetry. So there is now both a $Z^x_2$ bundle and an $SO(3)_s$ bundle on the ``physical" bulk manifold $X_3$. On the boundary $M$, $z_\alpha$ sees a combination of transition functions from $pin(2)_-$ and $SU(2)_s$. In fact, the transition functions for $z_{\alpha}$ live in $\left(pin(2)_- \times SU(2)_s\right)/Z_2$. Thus, the $pin(2)_-$ transition functions and the $SU(2)_s$ transition functions may not individually satisfy the cocycle condition, but the combination does. If we project our $pin(2)_-$ bundle $\xi_g$ to an $O(2)_g$ bundle $\tilde{\xi}_g$, and $SU(2)_s$ to an $SO(3)_s$ bundle $\xi_s$ then the resulting bundles satisfy 
\beq w_2[\tilde{\xi}_g] + w^2_1[\tilde{\xi}_g] = w_2[\xi_s] \label{pinspin}\eeq
 Indeed, the left and right-hand-sides are precisely the obstructions to lifting $\tilde{\xi}_g$ and $\xi_s$ to $pin(2)_-$ and $SU(2)_s$, respectively. Now, we extend the $\left(pin(2)_- \times SU(2)_s\right)/Z_2$ bundle from the surface $M$ to some $Y_3$ - the condition (\ref{pinspin}) continues to be satisfied on $Y_3$. This also automatically extends the $Z^x_2$ gauge field $x = w_1[\tilde{\xi}_g]$ to $Y_3$. Now, we want to check if (\ref{bulkbound3D1}) is still independent of the extension to $Y_3$. It suffices to compute $\int_{Y_3} x^3$ for $Y_3$ closed. We have, $x^3  = x (w_2[\tilde{\xi}_g] + w_2[\xi_s]) = \frac{dw_2[\tilde{\xi}_g]}{2} + x w_2[\xi_s]$, so $\int_{Y_3} x^3 = \int_{Y_3} x w_2[\xi_s]$, which generally does not vanish. However, there is an easy fix: we modify the action to be
\beq S_{bulk + bound} = S_{bound}[M] + \pi i \int_{X_3 \cup \bar{Y}_3} (x^3 + x w_2[\xi_s]) \label{bulkbound3D} \eeq
which now does not depend on the extension to $Y_3$ chosen. For $X_3$ closed, we recover (\ref{bulk3D}). The first term is a pure $Z^x_2$ anomaly, while the second term is a mixed $Z^x_2$, $SO(3)_s$ anomaly.

\section{CP$^1$ model in $2+1$D}
\label{app:CP2d}

In this appendix, we deduce the bulk action (\ref{bulk4D}), which matches the anomalies of 2+1D CP$^1$ model,
\beq L = |D_a z_{\alpha} |^2 + \frac{i}{2\pi} A \wedge da \label{eq:CP1xy2} \eeq
with $a$ - the dynamical gauge field and $A$ - a gauge field coupling to the flux current $\frac{1}{2\pi} db$. The symmetries of the CP$^1$ model we consider are $O(2)_\Phi = U(1)_{\Phi} \rtimes C$ and $SO(3)_s$ (see section \ref{sec:3Dpreamble}). 
 We denote the associated bundles by $\xi_{\Phi}$ and $\xi_s$.  Now, $SO(3)_s$ and $C$ act on the spinons $z$ in a projective manner (the $U(1)_\Phi$ group does not act on the spinons). Indeed, $C^2: z \to -z$. So, $C$ combines with the gauge group $u(1)_g$ to a group $pin(2)_-$. The overall transition functions seen by $z$ live in $(SU(2)_s \times pin(2)_-)/Z_2$. The transition functions of $SU(2)_s$ generally will satisfy the cocycle condition only up to a factor of $-1$, and so will the transition functions of $pin(2)_-$.  Let us project $pin(2)_-$ down to an $O(2)$ group that we call $O(2)_g$, and let the associated bundle be labeled by $\tilde{\xi}_g$. Then the obstruction to lifting $O(2)_g$ to $pin(2)_-$ must be exactly equal to $w_2(\xi_s)$. But the obstruction to lifting an $O(n)$ bundle to a $pin(n)_-$ bundle is $w_2 + w^2_1$.\cite{KirbyTaylor} So, we must have $w_2(\tilde{\xi}_g) + w^2_1(\tilde{\xi}_g) = w_2(\xi_s)$. We now extend the full  $O(2)_\Phi \times (SU(2)_s \times pin(2)_-)/Z_2$ bundle from our original 3-manifold $M$ to a 4-manifold $Y_4$, such that $\d Y_4 = M$ and define,
 \beq \frac{i}{2\pi} \int_{M} A \wedge da \equiv 2 \pi i \int_{Y_4} \frac{dA}{2\pi} \wedge \frac{da}{2\pi} \label{ExtBdb} \eeq
 $d A$ is the field strength of the $O(2)_\Phi$ bundle and $d a$ - the field strength of the $pin(2)_-$ bundle. Equivalently $2 d a$ is the field strength of the $O(2)_g$ bundle. We want to see if (\ref{ExtBdb}) is independent of the extension to $Y_4$, i.e. we want to find what values it takes for $Y_4$ closed. Since $w_1(\xi_\Phi) = w_1(\tilde{\xi}_g)$, we may combine the $O(2)_\Phi$ and $O(2)_g$ bundles into an $SO(4)$ bundle $\xi_\Phi \oplus \tilde{\xi}_g$. We claim, for closed $Y_4$,
 \beq  2 \pi i \int_{Y_4} \frac{dA}{2\pi} \wedge \frac{da}{2\pi} = \pi i w_4(\xi_\Phi \oplus \tilde{\xi}_g, Y_4) \label{Bdbw4}\eeq
Indeed, let's project $SO(4)$ to $SO(4)/Z_2 = SO(3)_L \times SO(3)_R$. $SO(2)$ rotations by angles $\alpha$, $\beta$ in $O(2)_\Phi$, $O(2)_g$ become rotations by $\alpha - \beta$ and $\alpha + \beta$ around (say) the $z$ axis in $SO(3)_L$ and $SO(3)_R$ respectively. The reflection ${\rm diag}(1,-1)$ performed simultaneously in $O(2)_\Phi$ and $O(2)_g$ becomes a simultaneous $\pi$ rotation around $y$ axis in  $SO(3)_L$ and $SO(3)_R$. Therefore, the $SO(3)_L$ and $SO(3)_R$ connections are (locally) $A^L = (A-2a) \left(\begin{array}{ccc} 0&-i &0\\i&0&0\\0&0&0\end{array}\right) $ and $A^R = (A+2a) \left(\begin{array}{ccc} 0&-i &0\\i&0&0\\0&0&0\end{array}\right) $. Now, for an $SO(4)$ bundle,
\beq w_4 = \frac{1}{4} (p_1[SO(3)_L] - p_1[SO(3)_R]) \quad {\rm (mod\,\, 2)}\eeq
(see Ref.~\cite{SO5gang}, Eq.~(141)). Here, $p_1$ is the Pontryagin number of an $SO(n)$ bundle, which has an integral formula: 
 \beq p_1[SO(n)] =
\frac{1}{2\cdot (2 \pi)^2} \int_{Y_4} {\rm tr}_{SO(n)} F\wedge 
F\eeq
So, 
\beq w_4[\xi_\Phi \oplus \tilde{\xi}_g, Y_4] = \frac{1}{4 (2\pi)^2} \int_{Y_4} \left((dA-2da)\wedge (dA-2da) - (dA+2da) \wedge (dA + 2da)\right) = - \frac{2}{(2\pi)^2} \int_{Y_4} dA\wedge da \nn\eeq
which proves (\ref{Bdbw4}). Next, let us use the Whitney sum formula,
\beq w_4[\xi_\Phi \oplus \xi_g] = w_2[\xi_\Phi] \cup w_2[\tilde{\xi}_g]\eeq
- all the other terms vanish, since $\xi_\Phi$ and $\tilde{\xi}_g$ are $O(2)$ bundles. Recalling $w_2(\tilde{\xi}_g) + w^2_1(\tilde{\xi}_g) = w_2(\xi_s)$ and $w_1(\tilde{\xi}_g) = w_1(\xi_\Phi)$, we have
\beq w_4[\xi_\Phi \oplus \tilde{\xi}_g] = w_2[\xi_\Phi] \cup (w_2[\xi_s] + w^2_1[\xi_\Phi])\eeq
Notice that any dependence on the gauge bundle $\tilde{\xi}_g$ has disappeared - the above expression only depends on the background gauge bundles of the global symmetries $O(2)_\Phi$ and $SO(3)_s$. This means that although (\ref{Bdbw4}) is dependent on the extension to $Y_4$, this dependence can be cancelled by thinking of the theory as living on the surface of a $3+1$D SPT. The bulk partition function of this SPT on a closed manifold $X_4$ is just,
\beq S_{bulk} = \pi i \int_{X_4}  w_2[\xi_\Phi] \cup (w_2[\xi_s] + w^2_1[\xi_\Phi]) \label{bulkCP1}\eeq
If $X_4$ has a boundary $M$ then we define,
\beq  S_{bulk + bound} = \int_{M} |D_a z_{\alpha} |^2 + 2 \pi i \int_{Y_4} \frac{dA}{2\pi} \wedge \frac{da}{2\pi} + \pi i \int_{X_4 \cup \bar{Y}_4}w_2[\xi_\Phi] \cup (w_2[\xi_s] + w^2_1[\xi_\Phi])\eeq
Now, any dependence on the extension to $Y_4$ is cancelled between the second and third term above. However, the action does depend on the values of the background $O(2)_\Phi \times SO(3)_s$ gauge fields on the ``physical" four-manifold $X_4$.

Note that we may also re-write $w_2[\xi_s] + w^2_1[\xi_\Phi] = w_2[\tilde{\xi}_s]$, where $\tilde{\xi}_s = \xi_s  \otimes \det(\xi_\Phi)$ is an $O(3)_s$ bundle derived from the original $SO(3)_s$ bundle $\xi_s$ by multiplying the transition functions by $-1$ whenever the rotation in $O(2)_\Phi$ is improper. 

Further note that as shown in Ref.~\onlinecite{ZoharRyan} we obtain the same anomaly by working with a different proposed formulation of the deconfined critical point - the $N_f= 2$ QCD$_3$ theory.\cite{SO5gang} Recall that the QCD$_3$ formulation has an anomalous global $SO(5)$ symmetry, and the anomaly is given by $S_{bulk} = \pi i w_4[\xi^5, X_4]$ where $\xi^5$ is an $SO(5)$ bundle. For the symmetries explicit in the CP$^1$ model, we $\xi^5 = \xi_\Phi \oplus \tilde{\xi_s}$ is a direct sum of $O(2)_\Phi$ bundle and $O(3)_s$ bundle. Using the Whitney formula, 
\beq w_4[\xi^5] = w_1[\xi_\Phi] w_3[\tilde{\xi}_s] + w_2[\xi_\Phi] w_2[\tilde{\xi}_s] \eeq
But $w_1[\xi_\Phi] = w_1[\tilde{\xi}_s]$ and $w_1 w_3 = \frac{dw_3}{2}$, so the first term is a total derivative and does not contribute to the bulk action. We then recover, $w_4[\xi^5] \to w_2[\xi_\Phi] w_2[\tilde{\xi}_s]$ in agreement with the computation in the CP$^1$ model. 

\subsection{Vanishing of anomaly}
\label{app:vanish}
We now show that the anomaly (\ref{bulkCP1}) vanishes for the symmetry appropriate to the honeycomb lattice and for the intrinsic symmetry appropriate to the $S = 1$ square lattice.

We begin with the honeycomb lattice. Here, the relevant subgroup of $O(2)_{\Phi}$ is $D_3$. We want to show that $w_2[\xi_\Phi] = 0$. Recall that $w_2$ is the obstruction to lifting an $O(n)$ bundle to a $pin(n)_+$ bundle.\cite{KirbyTaylor}  Let, $\pi: pin(2)_+ \to O(2)$ be the projection map. Now, $pin(2)_+ = O(2)$. Write, $O(2) = U(1)\rtimes Z_2$ with $Z_2$ generated by $C$. Then $\pi(u_\alpha) = u_{2 \alpha}$ and $\pi(C) = C$, where $u_\alpha$ is a rotation by $\alpha$ in $U(1)$. Furthermore, if we restrict $O(2)$ to a $D_3$ subgroup $\pi: D_3 \to D_3$ is an isomorphism. In fact, $\pi^2 = 1$. Thus, for any $D_3$ bundle we obtain a lift to $pin(2)_+$ simply by applying $\pi$ to the transition functions. Therefore, $w_2[D_3] = 0$.

Next, we proceed to the $S  = 1$ square lattice. Here, we want to show that the intrinsic anomaly vanishes. For this, we have to consider bundles associated with the microscopic symmetry group $(Z^x \times Z^y) \rtimes Z^{rot}_4$. Let $x$ be the generator of $Z^x$, $y$ the generator of $Z^y$ and $r$ the generator of $Z^{rot}_4$. The associated $O(2)_{\Phi}$ bundle is a $Z_2 \times Z_2$ bundle obtained via the projection $p: (Z^x \times Z^y) \rtimes Z^{rot}_4 \to Z_2 \times Z_2$, with $p(x) = C$, $p(y) = C$, $p(r) = u_{\pi}$. We can also form a $D_4$ representation $s: (Z^x \times Z^y) \rtimes Z^{rot}_4 \to D_4$, with $s(x) = C$, $s(y) = u_\pi C$ and $s(r) = u_{\pi/2}$. We then have the sequence $(Z^x \times Z^y)\rtimes Z^{rot}_4 \stackrel{s}{\to} D_4 \stackrel{\pi}{\to} Z_2 \times Z_2$, with $\pi: pin(2)_+ \to O(2)$ as before. Further, $\pi \circ s = p$. So to obtain a lift of $Z_2 \times Z_2$ to $D_4$, we simply apply $s$ to the parent $ (Z^x \times Z^y) \rtimes Z^{rot}_4$. Therefore, $w_2[\xi_{\Phi}] = 0$.

\section{Asymmetric Vortices}
\label{app:vort}

\begin{figure}[h]
    \centering
    \includegraphics{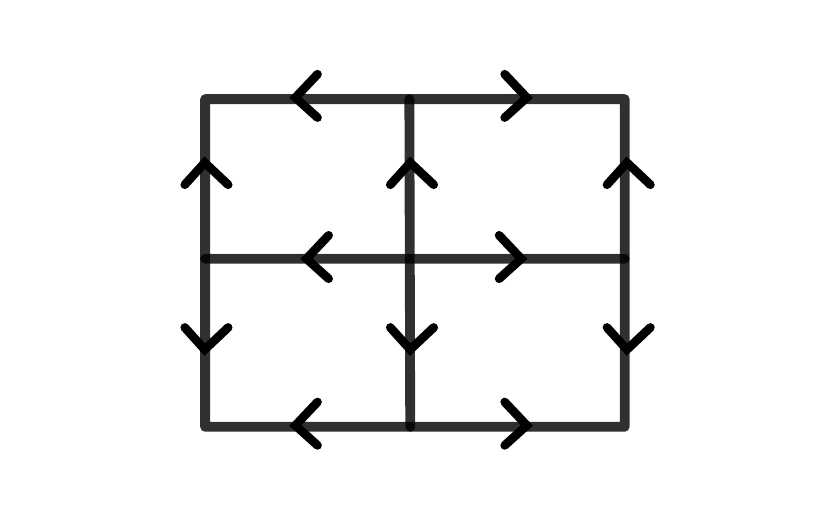}
    \caption{A unit cell for the branching structure of the usual VBS convention. Edges occupied with a dimer are considered part of a domain associated with the direction labeling that edge.}
    \label{fig:vbsbranching}
\end{figure}

In section \ref{sec:vort} we revisited the well-known fact that $Z^{rot}_4$ VBS vortices on the square lattice carry $S =1/2$ in their core. We emphasized that in general one needs to consider $Z^{rot}_4$ symmetric VBS vortices in order to reach this conclusion.
In our analysis, we defined a vortex as having four macroscopic VBS domains in a clock configuration. The details of the domain walls separating the domains did not affect the counting of the vortex winding.  In this appendix, we show that for the nearest neighbour dimer model there is an alternative way to define the vorticity by a closed line integral around a contour enclosing the vortex core, so that the vorticity does depend on the microscopic details of the domain walls. Further, with this definition, the vorticity is always equal to $N_A - N_B$, where $N_{A/B}$ is the number of ``dangling spins" on $A/B$ sites in the vortex core. This holds even when the vortex is not rotationally symmetric. Further, we use this definition of vorticity to make contact with the anomaly formula \eqref{bulkrot}: $S = \pi i \int_{X_4} \frac{d \gamma}{4} \cup w^s_2$.

\begin{figure}[h]
    \centering
    \includegraphics{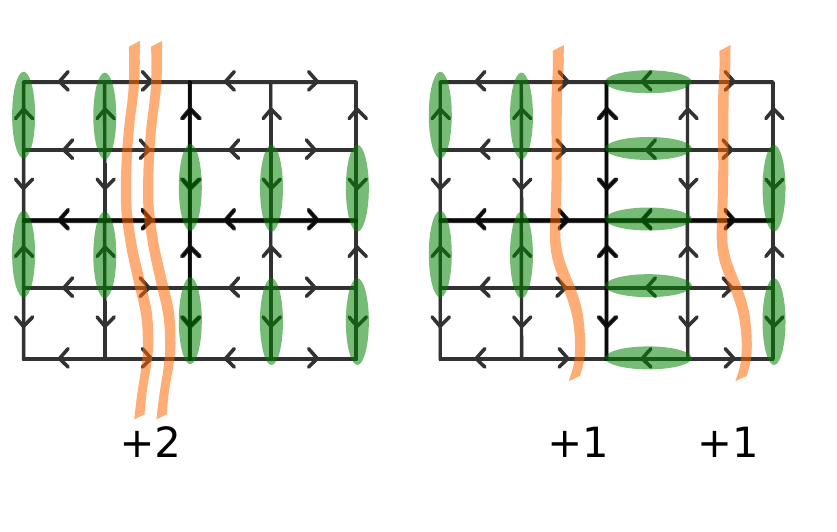}
    \caption{A domain wall accross which the direction of the VBS order (arrows highlighted by dimers) rotates by $\pi$ (left). This domain wall may be resolved as two $\pi/2$ domain walls (right), revealing that it is a counterclockwise (positive) $\pi$ rotation.}
    \label{fig:vbsbranching2}
\end{figure}

For a dimer configuration on the square lattice, we want to compute the ``vortex charge" $Q(U)$ of a region $U$. We assume that if any ``dangling" spins are present, they are away from the boundary $\d U$.  We define $Q(U) =  \frac{1}{4} \int_{\partial U} \gamma$, where $\gamma$ is a 1-cochain living on the links of the square lattice. This cochain is defined by counting VBS domain walls crossing the (oriented) contour $\partial U$ in the following way. First, we assign numbers $1, i, -1, -i$ to the links of the square lattice using a $2 \times 2$ unit cell as shown in figure \ref{fig:vbsbranching} ($1$ is represented by a right arrow, $i$ by an up arrow, $-1$ by a left arrow and $-i$ by a down arrow). For each site $j$, we define the VBS order parameter $V_j$ by the number on the dimer covering $j$ - this is the standard definition of the columnar dimer order parameter. Now, to define $\gamma$ on a link $j\mu$, $\mu = \hat{x}, \hat{y}$, we consider $\frac{V_{j+\mu}}{V_j}$. If $\frac{V_{j+\mu}}{V_j} = 1$, we set $\gamma_{j\mu} = 0$.  If $\frac{V_{j+\mu}}{V_j} \neq 1$,  the link crosses a VBS domain wall. For $\frac{V_{j+\mu}}{V_j} = \pm i$, this is a ``single" domain wall, and we assign $\gamma_{j \mu} = \pm 1$. For $\frac{V_{j+\mu}}{V_j} = -1$, we have a double domain wall and assign $\gamma_{j \mu} = \pm 2$. The sign can be determined by breaking up the double domain wall into two single domain walls, as demonstrated in Fig.~\ref{fig:vbsbranching2}. Using this procedure, we obtain the following general expression for the sign of $\gamma_{j\mu}$. Let $\lambda_{j x}=(-1)^{j_x}$, $\lambda_{j y} = i (-1)^{j_y}$ (so that $\lambda_{i \mu}$ coincides with the number we assigned to the corresponding link in Fig.~\ref{fig:vbsbranching}). If $\frac{V_{j+\mu}}{V_j} = -1$, $-\frac{\lambda_{j\mu}}{V_j} = \pm i$ and we define $\gamma_{j\mu} = \pm 2$.

\begin{figure}
    \centering
    \includegraphics{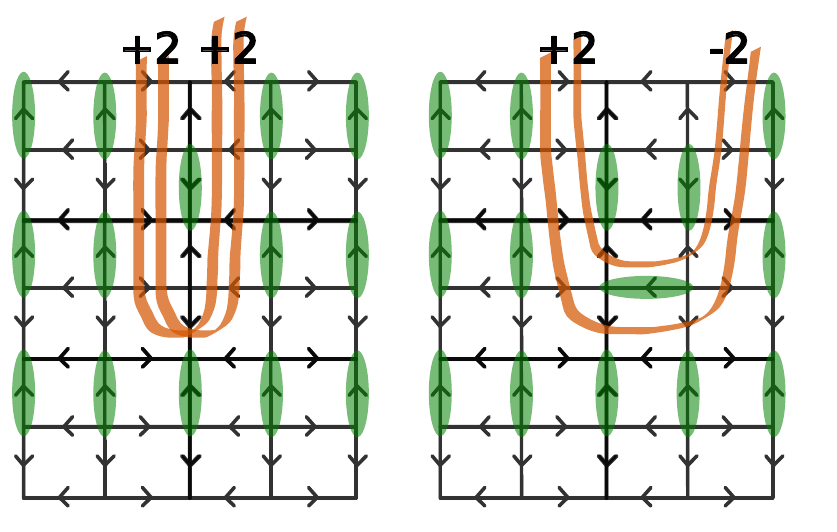}
    \caption{On the left, a $Q = -1$ VBS vortex (as measured by a counter-clockwise integration contour around the edge of the figure); note that the missing spin sits on the $B$ site, in agreement with $Q  =N_A-N_B$.  On the right, a VBS vortex and VBS anti-vortex sit side-by-side and their cores dimerize. The total winding number $Q = 0$.}
    \label{fig:vbsbranching3}
\end{figure}

A direct computation shows that away from ``dangling" spins $d \gamma  = 0$. Therefore, $Q(U)$ is invariant under deforming the boundary of $U$ (as long as we don't push the boundary through sites with dangling spins). One can also show that in terms of the two sublattices $A$ (those vertices   with all arrows incoming or all arrows outgoing) and $B$ (those vertices with two incoming arrows and two outgoing arrows), $Q(U)$ with a counterclockwise contour counts the number of unoccupied $A$ sites minus the number of unoccupied $B$ sites in $U$. Modulo 2, this just counts the number of unoccupied sites. Note that this identification works independent of the details of domain walls. For instance, in figure \ref{fig:vortsq1} top-left $Q  =1$, top-right $Q  =0$, and bottom $Q  =-3$, in agreement with $N_A - N_B$ (we take the unoccupied site in \ref{fig:vortsq1} top-left to be an $A$ site). Note, however, that there is no obvious way to extend this formula to more general dimer configurations (not just nearest neighbour). In particular, the integer nature of the invariant $Q$ is an artifact of only bipartite configurations being considered. Nevertheless, the formula for $Q$ is very reminiscent of the anomaly formula  \eqref{bulkrot}: $S = \pi i \int_{X_4} \frac{d \gamma}{4} \cup w^s_2$. Indeed, this formula indicates that in a spatial boundary slice $\Sigma$, $\int_{\Sigma} \frac{d\gamma}{4}$ (mod 2) tells us whether we have a projective $SO(3)$ representation or not. Identifying the cochain $\gamma$ extracted from the domain walls with the background $Z^{rot}_4$ gauge field $\gamma$ in the spirit of \cite{RyanDom}, we see a geometric confirmation of the anomaly formula.

We can also extend the definition of the vortex charge $Q$ to the honeycomb lattice. Here, we have three different Kekule VBS domains with $V = 1, e^{2 \pi i/3}, e^{4 \pi i /3}$. For a given link $(ij)$ we compute $\frac{V_j}{V_i}$. If $\frac{V_j}{V_i} = 1$, we assign $\gamma_{ij} = 0$ to the link. If $\frac{V_j}{V_i} = e^{\pm 2 \pi i/3}$, we assign $\gamma_{ij} = \pm 1$. Note, that in this case there are no double domain walls. It is again true that $Q  =N_A - N_B$. For instance, the vortex in figure Fig.~\ref{fig:vorth} left has $Q  =1$ and the vortex in figure Fig.~\ref{fig:vorth} right has $Q  =-2$, as required. 

\bibliography{AnDQCP}

\end{document}